\documentstyle[preprint,aps,prd]{revtex}
\input{epsf}
\tightenlines

\newcommand{\eq}{\begin{eqnarray}}
\newcommand{\en}{\end{eqnarray}}

\newcommand{\mathbold}[1]{\mbox{\boldmath $\bf#1$}}

\begin{document}

\draft
\title{Bound $q\bar q$ systems in the framework of different versions of
3D reductions of the Bethe-Salpeter equation}

\author{T.\ Kopaleishvili}

\address{HEPI, Tbilisi State University, University St. 9,
         380086 Tbilisi, Georgia}

\date{\today}

\maketitle

\begin{abstract}
Five different versions of the three-dimensional (3D) reduction of the 
Bethe-Salpeter (BS) equation in the  instantaneous approximation 
for kernel of BS equation for the two-fermion systems are formulated.
The normalization condition for the bound-state wave function in all
versions are derived.
Further, the 3D reduction of BS equation without instantaneous 
approximation for the kernel of BS equation is formulated in the 
quasi-potential approach. Except of the Salpeter version, other four 
versions have the correct one-body limit (Dirac equation) when mass of
one of constituent fermions tends to infinity. Application of these 
versions for investigation of the different properties of the $q\bar q$ 
bound systems are considered.
\end{abstract}


\section{Introduction}
After having firmly established the quark structure of mesons and baryons,
there naturally arises the question: how to describe the properties of
hadrons in terms of explicit quark and gluon degrees of freedom. The main
feature of QCD at low energy - confinement of quarks and gluons
into a colorless bound states - is still understood very little. For
this reason, one has to resort to various kinds of
QCD-inspired models. The simplest one is the so-called
constituent quark model, where quarks have a given ``constituent'' mass, and
the interactions between the
``constituent'' quarks within mesons $q\bar q$ and baryons $qqq$
are described by ``confining potentials'', growing to infinity at the
infinite quark separation. At the first stage,
this intuitive picture has been implemented within the
non-relativistic approach. Despite the evident success of the
non-relativistic potential model, it has been understood long time
ago, that one
has to include the relativistic effects, at least when describing
hadrons consisting of light $u,d,s$ quarks. Field-theoretical
Bethe-Salpeter (BS) equation provides a natural basis for a
relativistic generalization of the potential model, where both
light $u,d,s$ and the heavy quark $c,b$ bound states can be treated on the
equal footing. A more sophisticated approach is based on a coupled set of
Dyson-Schwinger (DS) and BS equations, that can be derived at QCD
level~\cite{Roberts,Tandy,Maris}. 
In such an approach, one uses a model gluon propagator that
through the solution of the DS equation leads to the quark propagator
which is an entire function in a complex $p^2$ plane and therefore
is believed to correspond to the confined quark. 
A full content of underlying QCD
symmetries which are important at low energy, can be consistently embedded
within this approach. In particular, the Goldstone bosons are
properly described, and in the limit of the vanishing quark masses, the
masses of Goldstone bosons obtained through the solution of the
coupled DS and BS equations, also vanish (Note that it is not
the case in the simple potential-type models with quarks having the constant 
``constituent'' mass). 

In the following, we
shall review the potential model based solely on the BS equation,
which is the subject of intensive investigations during last twenty years.

\section{Bethe-Salpeter equation for the two-fermion bound state}

To set up the notation, in this section we give a brief survey of the
covariant BS approach to the two-fermion (fermion-antifermion) bound states.
In order to derive the BS equation for the two-fermion bound state, we
consider the full 4-fermion Green function $G$ which in the momentum space
is given by
\eq\label{G}
G(p_1,p_2;p_1',p_2')=i^2\int&& dx_1dx_2dx_1'dx_2'\,\,
{\rm e}^{ip_1x_1+ip_2x_2-ip_1'x_1'-ip_2'x_2'}\,\times
\nonumber\\[2mm]
&&\times\langle 0
|T\psi_1(x_1)\psi_2(x_2)\bar\psi_1(x_1')\bar\psi_2(x_2')|0\rangle\, ,
\en
where, for simplicity, the fermions 1 and 2 are assumed to be
distinguishable, and the spinor indices are suppressed.

The Green function satisfies the BS equation in the momentum
space
\eq\label{BSG}
G=G_0+G_0KG=G_0+GKG_0\, ,
\en
where $G_0$ stands for the free 4-fermion Green function (the direct
product of two fermion propagators), and $K$ denotes the kernel of BS
equation, given by the sum of all two-particle irreducible Feynman graphs.

In the momentum space, it is convenient to define the center-of-mass (c.m.)
and relative 4-momenta according to the following relations (with arbitrary
$\alpha$ and $\beta$)\footnote{
We choose the system of units where $\hbar=c=1$. Any 4-vector has the
components $a=(a_0,{\bf a})$, and the metric is 
$g_{\mu\nu}={\rm diag}(1,-1,-1,-1)$}
\eq\label{c.m.}
&&P=p_1+p_2,\quad p=\beta p_1-\alpha p_2,\quad \alpha+\beta=1\, ,
\nonumber\\[2mm]
{\rm or}\quad\quad &&p_1=\alpha P+p,\quad p_2=\beta P-p\, .
\en
For the basis vectors in the momentum space, the following notation is used.
\eq\label{basicvectors}
|p_1\rangle\otimes|p_2\rangle=|p_1p_2\rangle=|Pp\rangle
=|P\rangle\otimes|p\rangle\, .
\en
These vectors satisfy the completeness and orthonormality conditions
\eq\label{completeness}
\int\,|p_i\rangle\, \frac{d^4p_i}{(2\pi)^4}\,\langle p_i|={\bf 1}\, \quad
{\rm for}~i=1,2\, ,\quad
\int\,|P\rangle\, \frac{d^4P}{(2\pi)^4}\,\langle P|={\bf 1}\, ,\quad
\int\,|p\rangle\, \frac{d^4p}{(2\pi)^4}\,\langle p|={\bf 1}\, ,
\en
\eq\label{orthonormality}
\langle p_i|p_j'\rangle=\delta_{ij}(2\pi)^4\delta^4(p_i-p_i')\, ,\quad
\langle P|P'\rangle=(2\pi)^4\delta^4(P-P')\, ,\quad
\langle p|p'\rangle=(2\pi)^4\delta^4(p-p')\, .
\en
In these notations, we can write
\eq\label{removal-c.m.}
\langle Pp|{\bf O}|P'p'\rangle=(2\pi)^4\delta^4(P-P')\,\,\biggl[
\langle p|{\bf O}(P)|p'\rangle\equiv {\bf O}(P;p,p')\biggr]\, ,
\quad {\bf O}=G,~G_0,~K\, .
\en
Further,
\eq\label{G0-1}
\langle p|G_0(P)|p'\rangle\equiv G_0(P;p,p')=(2\pi)^4\delta^4(p-p')\,
G_0(P;p)\, ,
\en
\eq\label{G0-2}
G_0(P;p)=S_1(p_1)\otimes S_2(p_2)=-(\not\! p_1+m_1)\otimes(\not\! p_2+m_2)
\,\, g_0(P;p)\, ,
\en
where $S_i(p_i)=i(\not\! p_i-m_i)^{-1}$ stands for the free fermion
propagator with the mass $m_i$, and the quantity $g_0(P;p)$ is defined
as follows
\eq\label{g_0}
g_0(P;p)=\frac{1}{p_1^2-m_1^2+i0}\,\frac{1}{p_2^2-m_2^2+i0}=
\frac{1}{p_{10}^2-w_1^2+i0}\,\frac{1}{p_{20}^2-w_2^2+i0}\, ,
\en
with $w_i=\sqrt{m_i^2+{\bf p}_i^2}$.

The stable bound state with the mass $M_B$ in quantum field theory is
described by the 1-particle state vector in the Fock space
\eq\label{boundstate}
\langle {\bf P}_B|{\bf P}_B'\rangle=(2\pi)^3\, 2w_B\,
\delta^3({\bf P}-{\bf P}')\, ,\quad\quad w_B=\sqrt{M_B^2+{\bf P}^2}\, .
\en
However, there is no interpolating field in the Lagrangian corresponding to
the bound state particle. The completeness condition of the Fock-state
vectors in the presence of bound states reads
\eq\label{completeness-B}
{\bf 1}=\int\,\, |{\bf P}_B\rangle\,\,\frac{d^3{\bf P}_B}{(2\pi)^3}\,\,
\langle {\bf P}_B|+\cdots
\en
where ellipses stand for the contributions of the states with elementary
particles and from the multi-particle scattering states.

Using the completeness condition~(\ref{completeness-B}), it is
straightforward to single out the bound-state contribution
in the Green function~(\ref{G})
when $P^2\rightarrow M_B^2$ (equivalently
$P_0^2\rightarrow w_B^2$).
The quantity $\langle p|G(P)|p'\rangle$
exhibits the pole behavior at this point
\eq\label{boundstatepole}
\langle p|G(P)|p'\rangle=i\,\frac{\langle p|\Phi_{{\bf P}_B}\rangle
\langle\bar\Phi_{{\bf P}_B}|p'\rangle}{P^2-M_B^2}+\langle p|R(P)|p'\rangle\, ,
\en
where $\langle p|R(P)|p'\rangle$ denotes the regular remainder of
$\langle p|G(P)|p'\rangle$ at the bound-state
pole that emerges from the contribution of other states in the sum over
Fock-space vectors. Further, $\langle p|\Phi_{{\bf P}_B}\rangle$ stands
for the BS wave function of the bound state
\eq\label{w.f.}
\langle p|\Phi_{{\bf P}_B}\rangle&\equiv&\Phi_{{\bf P}_B}(p)=
\int dx\,\,{\rm e}^{ipx}\,\,\langle 0|T\psi_1(\beta x)\psi_2(-\alpha x)
|{\bf P}_B\rangle\, ,
\nonumber\\[2mm]
\langle \bar\Phi_{{\bf P}_B}|p'\rangle&\equiv&\bar\Phi_{{\bf P}_B}(p')=
\int dx\,\,{\rm e}^{-ip'x}\,\,
\langle {\bf P}_B|T\bar\psi_1(\beta x)\bar\psi_2(-\alpha x)|0\rangle
=\Phi^\dagger_{{\bf P}_B}(p')\,\gamma_0^{(1)}\otimes\gamma_0^{(2)}
\en

The bound-state equation that can be derived for the state vector
$|\Phi_{{\bf P}_B}\rangle$ by substituting Eq.~(\ref{boundstatepole})
in the BS equation for the Green function~(\ref{BSG}), formally resembles the
nonrelativistic Schr\"{o}dinger equation for two fermions
\eq\label{BSwf-1}
G^{-1}(P_B)|\Phi_{{\bf P}_B}\rangle=0\, ,\quad\quad
\langle \bar\Psi_{{\bf P}_B}|G^{-1}(P_B)=0\, ,\quad
{\rm with}\quad G_0^{-1}(P)-G^{-1}(P)=K(P)\, ,
\en
or
\eq\label{BSwf-2}
|\Phi_{{\bf P}_B}\rangle=G_0(P_B)K(P_B)|\Phi_{{\bf P}_B}\rangle\, ,\quad\quad
\langle \bar\Phi_{{\bf P}_B}|=\langle \bar\Phi_{{\bf P}_B}|K(P_B)G_0(P_B)\, .
\en
Here $P_B=(w_B,{\bf P}_B)$. Explicitly, in the momentum space, we arrive
at the following equation for the bound-state wave function~\cite{BS}
\eq\label{BSwf-3}
\Phi_{{\bf P}_B}(p)&=&G_0(P;p)\,\int\,\frac{d^4p'}{(2\pi)^4}\,\,
K(P;p,p')\Phi_{{\bf P}_B}(p)\, ,
\nonumber\\[2mm]
\bar\Phi_{{\bf P}_B}(p)&=&\int\,\frac{d^4p'}{(2\pi)^4}\,\,
\bar\Phi_{{\bf P}_B}(p')K(P;p',p)G_0(P;p)\, .
\en
This equation should be solved in order to obtain the mass $M_B$ of the bound
state. It is obvious that both equations: for $\Phi_{{\bf P}_B}(p)$, and for
its conjugate $\bar\Phi_{{\bf P}_B}(p)$, lead to the same bound-state
spectrum.

Next, we derive the normalization condition for the BS wave function.
To this end, it is useful to start from the following identity
\eq\label{identity}
G(P)G^{-1}(P)G(P)=G(P)\Rightarrow G(P)(G_0^{-1}(P)-K(P))G(P)=G(P)\, .
\en
If $P^2$ is close to $M_B^2$, one can neglect the contribution from $R(P)$
in Eq.~(\ref{boundstatepole}). We substitute the latter into
Eq.~(\ref{identity}), and perform the integration along the closed contour
$C$ that encircles only the bound-state pole at $P_0=w_B$,
in the complex $P_0$ plane.
\eq\label{norm-derivation}
&&i\int_C|\Phi_{{\bf P}_B}\rangle\langle\bar\Phi_{{\bf P}_B}|\,\,
\frac{(G_0^{-1}(P)-K(P))\, dP_0}{(P_0+w_B-i0)^2(P_0-w_B+i0)^2}\,\,
|\Phi_{{\bf P}_B}\rangle\langle\bar\Phi_{{\bf P}_B}|
\nonumber\\[2mm]
&=&\int_C|\Phi_{{\bf P}_B}\rangle\,\,\frac{dP_0}{(P_0+w_B-i0)(P_0-w_B+i0)}\,\,
\langle\bar\Phi_{{\bf P}_B}|\, .
\en

From the Cauchy's theorem, one has
\eq\label{Cauchy}
\int_C\,\frac{f(z)\, dz}{(z-z_S)^n}=\pm 2\pi i\,\,
\frac{(-)^{n-1}}{(n-1)!}\,\frac{d^{n-1}}{dz^{n-1}}\,f(z)\biggr|_{z=z_S}
\, ,
\en
where the function $f(z)$ is analytic inside the contour $C$, and the choice
of the $\pm$ sign depends on whether one integrates counterclockwise ($+$)
or clockwise ($-$) along the contour.
With the use of the above formula, from Eq.~(\ref{norm-derivation}) one
readily obtains the normalization condition for the BS wave function
\eq\label{normalization}
i\int\,\frac{d^4p}{(2\pi)^4}\,\frac{d^4p'}{(2\pi)^4}\,\,
\bar\Phi_{{\bf P}_B}(p)\,\,\biggl[\frac{\partial}{\partial P_0}
(G_0^{-1}(P;p,p')-K(P;p,p'))\biggr]_{P_0=w_B}\Phi_{{\bf P}_B}(p')=2w_B
\, .
\en
The equations~(\ref{BSwf-3}), together with the normalization
condition~(\ref{normalization}), completely determine the BS mass spectrum
and the BS wave function.

At the end of this section, we shall consider in some detail the spin content
of the BS wave function. In particular, we shall demonstrate that
one can rewrite this equation in terms of ``fermion-antifermion'' rather than
``two fermion'' wave function.

We work with the following representation of Dirac $\gamma$-matrices
\eq\label{gamma}
\gamma^0=\gamma_0=\pmatrix{I&0\cr 0&-I}\, ,\quad\quad
\mathbold{\gamma}=\gamma^0\biggl[\,\,
\mathbold{\alpha}=\pmatrix{0&\mathbold{\sigma}\cr \mathbold{\sigma}&0}\,
\biggr]=-\mathbold{\alpha}\gamma^0\, .
\en

The free two-fermion Green function given by Eqs.~(\ref{G0-2}), (\ref{g_0}),
can be written as
\eq\label{G0-2X2}
G_0(P;p)=\pmatrix{G_{aa}(p_1)&G_{ab}(p_1)\cr G_{ba}(p_1)&G_{bb}(p_1)}\otimes
\pmatrix{G_{aa}(p_2)&G_{ab}(p_2)\cr G_{ba}(p_2)&G_{bb}(p_2)}\, ,
\en
where $G_{uv}(p_i),~u,v=a,b$ is the $2\times 2$ matrix (operator) in the
spin space of the $i$-th particle. Further, the BS wave function of
the two-fermion system can be written as a column
\eq\label{wf-2X2}
\Phi_{{\bf P}_B}(p)=\pmatrix{\Phi_{aa}(P;p)\cr \Phi_{ab}(P;p)\cr
  \Phi_{ba}(P;p)\cr \Phi_{bb}(P;p)}\, ,
\en
where, again, the components $\Phi_{uv}(P;p),~u,v=a,b$ are the $2\times 2$
matrices in the spin space of two fermions.

Now, it is straightforward to ensure that the BS equation~(\ref{BSwf-3})
can be rewritten in terms of ``fermion-antifermion'' wave function
$\Psi_{{\bf P}_B}(p)$
\eq\label{BS-C}
\Psi_{{\bf P}_B}(p)=S^{(1)}(p_1)\int\frac{d^4p'}{(2\pi)^4}\,\,
K(P;p,p')\Psi_{{\bf P}_B}(p')\, S^{(2)}(-p_2)\, .
\en
The wave functions $\Psi_{{\bf P}_B}(p)$ and $\Phi_{{\bf P}_B}(p)$ are
related by (see~\cite{Itzykson})
\eq\label{Psi-C}
\Psi_{{\bf P}_B}(p)=\pmatrix{\Phi_{aa}(P;p)&\Phi_{ab}(P;p)\cr
\Phi_{ba}(P;p)&\Phi_{bb}(P;p)}\, C=
-i\pmatrix{\Phi_{ab}(P;p)\sigma_2&\Phi_{aa}(P;p)\sigma_2\cr
\Phi_{bb}(P;p)\sigma_2&\Phi_{ba}(P;p)\sigma_2}\, ,
\en
where
\eq\label{C}
C=i\gamma^2\gamma^0=\pmatrix{0&-i\sigma_2\cr -i\sigma_2&0}
\en
denotes the charge conjugation matrix.

\section{Three-dimensional reductions of the BS equation}
\setcounter{equation}{0}

One of the reasons why the three-dimensional (3D) reduction of the BS
equation is necessary, is the absence of the usual quantum-mechanical
probability interpretation for the wave function $\Phi_{{\bf P}_B}(p)$ due to
the dependence of the latter on the $0$-th component of the relative
4-momentum. Further, in the presence of the confining interactions,
it is extremely difficult to construct a ``reasonable''
kernel $K$ in four dimensions that describes these interactions - we are not
aware of any, completely successful attempt. On the other hand, the concept
of static (3D) confining kernels that corresponds to an intuitively clear
picture of infinitely rising potentials in the coordinate space,
has been extremely useful in many semi-phenomenological applications to study,
e.g. the characteristics of heavy quarkonia, etc.

For this reason, below we shall mainly consider the static BS kernels
(i.e. the kernels which do not depend on the c.m. momentum $P$ and the $0$-th
components of the relative momenta $p_0,~p_0'$)
\eq\label{static}
K(P;p,p')\rightarrow K_{st}({\bf p},{\bf p}')\equiv -iV({\bf p},{\bf p}')\, .
\en
In this approximation, there are still different versions of the 3D equations
for the bound-state wave function. Below, we shall consider these versions in
detail.

\subsection{The Salpeter equation~[6]}

In the approximation~(\ref{static}), from Eq.~(\ref{BSwf-3}) it is
straightforwardly obtained
\eq\label{Sal-0}
\Phi_{{\bf P}_B}(p)&=& G_0(P;p)\,
\int\frac{d^3{\bf p}'}{(2\pi)^3}\, K_{st}({\bf p},{\bf p}')\,
\tilde \Phi_{{\bf P}_B}({\bf p}')\, ,
\nonumber\\[2mm]
{\bar \Phi}_{{\bf P}_B}(p)&=&\int\frac{d^3{\bf p}'}{(2\pi)^3}\,
\tilde{\bar \Phi}_{{\bf P}_B}({\bf p}')\,K_{st}({\bf p}',{\bf p})\,
G_0(P;p)\, ,
\en
and
\eq\label{Sal-1}
\tilde \Phi_{{\bf P}_B}({\bf p})&=&\tilde G_0(P;{\bf p})\,
\int\frac{d^3{\bf p}'}{(2\pi)^3}\, V({\bf p},{\bf p}')\,
\tilde \Phi_{{\bf P}_B}({\bf p}')\, ,
\nonumber\\[2mm]
\tilde{\bar \Phi}_{{\bf P}_B}({\bf p})&=&\int\frac{d^3{\bf p}'}{(2\pi)^3}\,
\tilde{\bar \Phi}_{{\bf P}_B}({\bf p}')\,V({\bf p}',{\bf p})\,
\tilde G_0(P;{\bf p})\, ,
\en
where
\eq\label{equal-time}
\tilde \Phi_{{\bf P}_B}({\bf p})=\int\frac{dp_0}{2\pi}\,
\Phi_{{\bf P}_B}(p)\, ,\quad
\tilde {\bar \Phi}_{{\bf P}_B}({\bf p})=\int\frac{dp_0}{2\pi}\,
\bar\Phi_{{\bf P}_B}(p)\, ,\quad
\tilde G_0(P;{\bf p})=\int\frac{dp_0}{2\pi i}\, G_0(P;p)\, .
\en
At the next step, we introduce the projection operators
\eq\label{projection}
\Lambda^{(\pm)}_i({\bf p}_i)=\frac{w_i\pm h_i({\bf p}_i)}{2w_i}\, ,\quad\quad
h_i({\bf p}_i)=\mathbold{\alpha}^{(i)}{\bf p}_i
+m_i\gamma^{(i)}_0\, ,\quad\quad i=1,2\, ,
\en

with the properties
\eq\label{projection-properties}
\sum_{\alpha_i=\pm}\Lambda_i^{(\alpha_i)}=1\, ,\quad\quad
\Lambda_i^{(\alpha_i)}\Lambda_i^{(\alpha_i')}=\delta_{\alpha_i\alpha_i'}
\Lambda_i^{(\alpha_i)}\, ,\quad\quad
h_i({\bf p}_i)\Lambda_i^{(\pm)}=\pm w_i\Lambda_i^{(\pm)}\, .
\en
With the use of the following identity
\eq\label{projection-identity}
\not\! p_i+m_i=\biggl\{ (p_{i0}+w_i)\Lambda_i^{(+)}({\bf p}_i)
+(p_{i0}-w_i)\Lambda_i^{(-)}({\bf p}_i)\biggr\} \gamma_0^{(i)}\, ,
\en
it is straightforward to obtain
\eq\label{equal-time-Green}
\tilde G_0(P;{\bf p})&=&\biggl\{
\frac{\Lambda_{12}^{(++)}({\bf p}_1,{\bf p}_2)}{P_0-w_1-w_2+i0}
-\frac{\Lambda_{12}^{(--)}({\bf p}_1,{\bf p}_2)}{P_0+w_1+w_2}\biggr\}
\gamma_0^{(1)}\otimes\gamma_0^{(2)}
\nonumber\\[2mm]
&=& \bigl[\, P_0-h_1({\bf p}_1)-h_2({\bf p}_2)\, \bigr]^{-1}
(\Lambda_{12}^{(++)}({\bf p}_1,{\bf p}_2)
-\Lambda_{12}^{(--)}({\bf p}_1,{\bf p}_2))
\,\gamma_0^{(1)}\otimes\gamma_0^{(2)}\, ,
\en
where $\Lambda_{12}^{(\alpha_1\alpha_2)}({\bf p}_1,{\bf p}_2)=
\Lambda_1^{(\alpha_1)}({\bf p}_1)\otimes\Lambda_2^{(\alpha_2)}({\bf p}_2)$.

Now the Salpeter equation~(\ref{Sal-1}) in the c.m. frame
(${\bf P}_B={\bf 0}$) can be written as
\eq\label{Sal-2}
\bigl[\, P_0-h_1({\bf p}_1)-h_2({\bf p}_2)\, \bigr]\tilde\Phi_{M_B}({\bf p})
=\Pi({\bf p})\,\int\frac{d^3{\bf p}'}{(2\pi)^3}\,
\gamma_0^{(1)}\otimes\gamma_0^{(2)}\, V({\bf p},{\bf p}')\,
\tilde\Phi_{M_B}({\bf p}')\, ,
\en
where
\eq\label{Pi}
\Pi({\bf p})=(\Lambda_{12}^{(++)}({\bf p}_1,{\bf p}_2)
-\Lambda_{12}^{(--)}({\bf p}_1,{\bf p}_2))=
\frac{h_1({\bf p})}{2w_1}+\frac{h_2(-{\bf p})}{2w_2}\, .
\en
Introducing the ``frequency components'' of the wave function according to
\eq\label{frequency}
\tilde\Phi_{{\bf P}_B}({\bf p})=\sum_{\alpha_1\alpha_2}
\tilde\Phi_{{\bf P}_B}^{(\alpha_1\alpha_2)}({\bf p})\, ,\quad\quad
\tilde\Phi_{{\bf P}_B}^{(\alpha_1\alpha_2)}({\bf p})=
\Lambda_{12}^{(\alpha_1\alpha_2)}({\bf p}_1,{\bf p}_2)
\,\tilde\Phi_{{\bf P}_B}({\bf p})\, ,
\en
the Eq.~(\ref{Sal-2}) can be reduced to the following system of equations
\eq\label{Sal-3}
\bigl[ M_B\mp (w_1+w_2)\bigr]
\tilde\Phi_{M_B}^{(\pm\pm)}({\bf p})
=\pm\Lambda_{12}^{(\pm\pm)}({\bf p},-{\bf p})
\,\gamma_0^{(1)}\otimes\gamma_0^{(2)}\,
\int\frac{d^3{\bf p}'}{(2\pi)^3}\, V({\bf p},{\bf p}')\,
\tilde\Phi_{M_B}({\bf p}')\, ,
\en
with additional conditions
\eq\label{additional}
\tilde\Phi_{M_B}^{(\pm\mp)}({\bf p})=0, ,\quad\quad
\tilde\Phi_{M_B}({\bf p})
=\tilde\Phi_{M_B}^{(++)}({\bf p})
+\tilde\Phi_{M_B}^{(--)}({\bf p})\, .
\en

The normalization condition can be readily obtained from
Eq.~(\ref{normalization}) by using the approximation~(\ref{static}) for the
kernel, the relation between 4D and 3D wave functions (\ref{Sal-0}), and the
decomposition of the wave function~(\ref{frequency}), (\ref{additional})
\eq\label{normalization-Salpeter}
\int\frac{d^3{\bf p}}{(2\pi)^3}\,
\bigl\{\, |\tilde\Phi_{M_B}^{(++)}({\bf p})|^2
-|\tilde\Phi_{M_B}^{(--)}({\bf p})|^2\,\bigr\}=2M_B\, .
\en

Note that the wave function $\tilde\Phi_{M_B}({\bf p})$ can be represented in
a form analogous to~(\ref{wf-2X2})
\eq\label{Salpeter-2X2}
\tilde\Phi_{M_B}({\bf p})=
\pmatrix{\tilde\Phi_{aa}({\bf p})\cr
         \tilde\Phi_{ab}({\bf p})\cr
         \tilde\Phi_{ba}({\bf p})\cr
         \tilde\Phi_{bb}({\bf p})}\, .
\en
The constraints~(\ref{additional}) can be considered as equations for the
components  $\tilde\Phi_{ab}({\bf p})$ and $\tilde\Phi_{ba}({\bf p})$. The
solution of these equations gives
\eq\label{Salpeter-ab}
\tilde\Phi_{ab}&=&(m_1w_2+m_2w_1)^{-1}\,\bigl\{
 w_1(\mathbold{\sigma}^{(2)}{\bf p}_2) \tilde\Phi_{aa}
-w_2(\mathbold{\sigma}^{(1)}{\bf p}_1) \tilde\Phi_{bb}\bigr\}\, ,
\nonumber\\[2mm]
\tilde\Phi_{ba}&=&(m_1w_2+m_2w_1)^{-1}\,\bigl\{
 w_2(\mathbold{\sigma}^{(1)}{\bf p}_1) \tilde\Phi_{aa}
-w_1(\mathbold{\sigma}^{(2)}{\bf p}_2) \tilde\Phi_{bb})\bigr\}\, .
\en
For the ``frequency components''
$\tilde\Phi^{(\alpha_1\alpha_2)}_{xy}
=\Lambda_{12}^{(\alpha_1\alpha_2)}({\bf p})\tilde\Phi_{xy}({\bf p}),~~
x,y=a,b$,
we obtain the following relations
\eq\label{Salpeter-frequency}
\tilde\Phi^{(\pm\pm)}_{aa}&=&\pm(2(m_1w_2+m_2w_1))^{-1}\,\bigl\{
(w_1\pm m_1)(w_2\pm m_2)\tilde\Phi_{aa}
-(\mathbold{\sigma}^{(1)}{\bf p}_1)(\mathbold{\sigma}^{(2)}{\bf p}_2)\tilde\Phi_{bb}\bigr\}\, ,
\nonumber\\[2mm]
\tilde\Phi^{(\pm\pm)}_{bb}&=&\pm(2(m_1w_2+m_2w_1))^{-1}\,\bigl\{
(\mathbold{\sigma}^{(1)}{\bf p}_1)(\mathbold{\sigma}^{(2)}{\bf p}_2)\tilde\Phi_{aa}
-(w_1\mp m_1)(w_2\mp m_2)\tilde\Phi_{bb}\bigr\}\, ,
\nonumber\\[2mm]
\tilde\Phi^{(\pm\pm)}_{ab}&=&\pm(2(m_1w_2+m_2w_1))^{-1}\,\bigl\{
(w_1\pm m_1)(\mathbold{\sigma}^{(2)}{\bf p}_2)\tilde\Phi_{aa}
-(w_2\mp m_2)(\mathbold{\sigma}^{(1)}{\bf p}_1)\tilde\Phi_{bb}\bigr\}\, ,
\nonumber\\[2mm]
\tilde\Phi^{(\pm\pm)}_{ba}&=&\pm(2(m_1w_2+m_2w_1))^{-1}\,\bigl\{
(w_2\pm m_2)(\mathbold{\sigma}^{(1)}{\bf p}_1)\tilde\Phi_{aa}
-(w_1\mp m_1)(\mathbold{\sigma}^{(2)}{\bf p}_2)\tilde\Phi_{bb}\bigr\}\, .
\en

The normalization condition~(\ref{normalization-Salpeter}) can be rewritten
as
\eq\label{normalization-ab}
\int\frac{d^3{\bf p}}{(2\pi)^3}\,\,\frac{2w_1w_2}{m_1w_2+m_2w_1}\,\,
\bigl\{\, |\tilde\Phi_{aa}({\bf p})|^2-|\tilde\Phi_{bb}({\bf p})|^2\,
\bigr\}=2M_B\, .
\en

At the end of this subsection, we shall consider the existence of the
one-body limit in the Salpeter equation. From the physical point of view, it
is clear that if the mass of one of the particles in the two-particle bound
state tends to infinity, the equation for the wave function should reduce
to Dirac equation for the light particle with a given interaction potential.
Let us check this property for the Salpeter equation assuming, e.g. that the
mass of the first particle tends to infinity. In this limit,
\eq\label{limit}
m_1\rightarrow\infty\quad\quad \Rightarrow\quad\quad w_1\rightarrow
m_1,\quad\gamma_0^{(1)}\rightarrow 1, \quad h_1\rightarrow m_1\, .
\en

Then, the Salpeter equation for the bound-state vector
$|\tilde\Phi_{M_B}\rangle$ is reduced to ($E_2\equiv M_B-m_1$)
\eq\label{Salpeter-1-body}
\bigl(\, E_2-h_2\,\bigr)\,|\tilde\Phi_{E_2}\rangle
=\frac{1}{2}\,\biggl(1+\frac{h_2}{w_2}\biggr)\,\gamma_0^{(2)}\,
|\tilde\Phi_{E_2}\rangle\, .
\en
Due to the presence of the prefactor
$\frac{1}{2}\,\biggl(1+\frac{h_2}{w_2}\biggr)$, this equation differs from
the Dirac equation for the particle 2 moving in the potential $V$ - that is,
the Salpeter equation does not possess the correct one-body limit.

Now there arises an important problem to solve. We are willing to obtain the
3D reduction of the BS equation in the static approximation,
that correctly reproduces the dynamics of the system in the one-body
limit - this property might be important for the description,
e.g. the heavy-light $q\bar q$ bound states.

Below, we shall consider several versions of the 3D reduction procedure,
which lead to the correct one-body limit.

\subsection{The Gross equation [7]}

In the derivation of the Gross equation, first we assign $\alpha=0$ and
$\beta=1$, in the definition of the c.m. and relative momentum
variables~(\ref{c.m.}). Physically, this means that the whole c.m. momentum
is carried by the particle 2. The free Green function has the form
\eq\label{G0-Gross}
G_0(P;p)&=&-(\not\! p+m_1)\otimes (\not\! P-\not\! p+m_2)\,\, g_0(P;p)\, ,
\nonumber\\[2mm]
g_0(P;p)&=&\frac{1}{p^2-m_1^2+i0}\,\,\frac{1}{(P-p)^2-m_2^2+i0}\, .
\en
The first propagator can be rewritten as
\eq\label{principal-Gross}
\frac{1}{p^2-m_1^2+i0}=P\,\frac{1}{p^2-m_1^2}-i\pi\delta(p^2-m_1^2)\, ,
\en
where the symbol $P$ stands for the principal-value prescription.
The approximation that leads to the Gross equation, consists in the
substitution
\eq\label{approximation-Gross}
\frac{1}{p^2-m_1^2+i0}\,\, \Rightarrow\,\, -2\pi i\,\frac{\delta(p_0-w_1)}{2w_1}\, .
\en
This approximation is called the ``spectator approximation''.
Note that, in this approximation it is not only the principal-value
term in the propagator of the first particle that is neglected,
but also the term containing $\delta(p_0+w_1)$ that emerges from
$\delta(p^2-m_1^2)$. Consequently, in this approximation the particle 1
always stays on its mass shell defined by the equation $p_0=w_1$.
In a result of this approximation, the free Green function in the c.m.
frame ($P_\mu=(P_0,{\bf 0})$) can be rewritten in the following form
\eq\label{G0-Gr}
&&G_0^{\rm GR}(P_0;p)=2\pi i\delta(p_0-w_1)\tilde G_0^{\rm GR}(P_0;{\bf p})
\nonumber\\[2mm]
&=&2\pi i\delta(p_0-w_1) \biggl\{
 \frac{\Lambda_1^{(+)}({\bf p})\otimes\Lambda_2^{(+)}(-{\bf p})}{P_0-w_1-w_2+i0}
+\frac{\Lambda_1^{(+)}({\bf p})\otimes\Lambda_2^{(-)}(-{\bf p})}{P_0-w_1+w_2+i0}\biggr\}\,
\gamma_0^{(1)}\otimes\gamma_0^{(2)}\, ,
\en
where the functions $G_0^{\rm GR}(P_0;p)$ and 
$\tilde G_0^{\rm GR}(P_0;{\bf p})$ are related by Eq.~(\ref{equal-time}).

After substituting Eq.~(\ref{G0-Gr}) into~(\ref{BSwf-2}) and integrating
over the variable $p_0$, in the c.m. frame (now $P^\mu=(M_B,{\bf 0})$)
we arrive at the Gross equation for the 3D bound-state
wave function
\eq\label{Gross-wf-1}
\tilde\Phi_{M_B}({\bf p})&=&\tilde G_0^{\rm GR}(M_B;{\bf p})\,\,
\int\frac{d^3{\bf p}'}{(2\pi)^3}\,\, V({\bf p},{\bf p}')\,\tilde\Phi_{M_B}({\bf p}')\, ,
\nonumber\\[2mm]
\tilde{\bar\Phi}_{M_B}({\bf p})&=&
\int\frac{d^3{\bf p}'}{(2\pi)^3}\,\,\tilde{\bar\Phi}_{M_B}({\bf p}')\, V({\bf p}',{\bf p})\,\,
\tilde G_0^{\rm GR}(M_B;{\bf p})\, .
\en

Now, using again~(\ref{G0-Gr}) together with~(\ref{projection-properties}),
we arrive at
\eq\label{Gross-wf-2}
\bigl[ M_B-h_1({\bf p})-h_2(-{\bf p})\bigr]\,\,\tilde{\bar\Phi}_{M_B}({\bf p})=
\frac{1}{2}\biggl( 1+\frac{h_1}{w_1}\biggr)\gamma_0^{(1)}\otimes\gamma_0^{(2)}
\int\frac{d^3{\bf p}'}{(2\pi)^3}\,\, V({\bf p},{\bf p}')\,\tilde\Phi_{M_B}({\bf p}')\, ,
\en
which has the correct one-body limit when $m_1\rightarrow\infty$.

The normalization condition for the 3D wave function that satisfies the Gross equation,
can not be obtained in a standard manner, by using Eq.~(\ref{normalization}).
In order to demonstrate this,
note that according to Eqs.~(\ref{BSwf-2}), (\ref{G0-Gr}) and (\ref{Gross-wf-1}), 4D and 3D
wave functions are related by
\eq\label{Gross-interconnection}
\Phi_{M_B}(p)=2\pi\delta(p_0-w_1)\, \tilde\Phi_{M_B}({\bf p})\, ,\quad\quad
\bar\Phi_{M_B}(p)=2\pi\delta(p_0-w_1)\, \tilde{\bar\Phi}_{M_B}({\bf p})\, .
\en

Now if in the normalization condition~(\ref{normalization})
with the static kernel~(\ref{static}),
the relation~(\ref{Gross-interconnection}) between the 4D and
3D wave functions is substituted, one arrives at the ill-defined expression containing
the product of $\delta$-functions with the same argument.
For this reason, instead of the rigorous derivation, from the analogy
with the Salpeter equation, one merely assumes that the solutions of the Gross
equation satisfy the following normalization condition
\eq\label{Gross-normalization}
\int\frac{d^3{\bf p}}{(2\pi)^3}\,
\bigl\{ |\tilde\Phi_{M_B}^{(++)}({\bf p})|^2+ |\tilde\Phi_{M_B}^{(+-)}({\bf p})|^2\bigr\}
=2M_B\, .
\en

\subsection{The Mandelzweig-Wallace equation [8]}

In the derivation of the Mandelzweig-Wallace (MW) equation,
the parameters $\alpha$ and $\beta$ in the expression
of the c.m. and relative momenta~(\ref{c.m.}) are defined according to
Wightmann and Garding
\eq\label{WG}
\alpha=\alpha(s)=\frac{s+m_1^2-m_2^2}{2s}\, ,\quad\quad
\beta=\beta(s)=\frac{s-m_1^2+m_2^2}{2s}\, ,\quad\quad
s=P^2=P_0^2-{\bf P}^2\, .
\en

In the c.m. frame, from Eqs.~(\ref{c.m.}) and (\ref{WG}) it follows
\eq\label{WG-1}
&&p_1=(E_1+p_0,{\bf p})\, ,\quad
p_2=(E_2-p_0,-{\bf p})\, ,\quad
E_1=\frac{M_B^2+m_1^2-m_2^2}{2M_B}\, ,\quad
E_2=\frac{M_B^2-m_1^2+m_2^2}{2M_B}\,
\nonumber\\[2mm]
&&E_1+E_2=M_B\, ,\quad E_1-E_2=\frac{m_1^2-m_2^2}{M_B}\, .
\en

Further, we define in the c.m. frame
\eq\label{G0-MW}
\tilde G_0^{\rm MW}(M_B;{\bf p})=\int\frac{dp_0}{2\pi i}\,
\bigl[\, G_0(p_1,p_2)+G_0(p_1,p_2^{\rm cr})\, \bigr]\, ,\quad\quad
p_2^{\rm cr}=(E_2+p_0,-{\bf p})\, ,
\en
where $G_0(p_1,p_2)$ is given by Eqs.~(\ref{G0-1}), (\ref{G0-2}) and (\ref{g_0}).
After integrating over $p_0$, we obtain
\eq\label{G0-MW-1}
\tilde G_0^{\rm MW}(M_B;{\bf p})&=&\biggl\{
 \frac{\Lambda_{12}^{++}({\bf p},-{\bf p})}{E_1+E_2-w_1-w_2+i0}
+\frac{\Lambda_{12}^{+-}({\bf p},-{\bf p})}{-E_1+E_2+w_1+w_2}
\nonumber\\[2mm]
&+&\frac{\Lambda_{12}^{-+}({\bf p},-{\bf p})}{E_1-E_2+w_1+w_2}
-\frac{\Lambda_{12}^{--}({\bf p},-{\bf p})}{E_1+E_2+w_1+w_2}
\biggr\}\, \gamma_0^{(1)}\otimes\gamma_0^{(2)}\, .
\en
The MW equation is obtained from the BS equation in the
static approximation, by using the combination
$G_0(p_1,p_2)+G_0(p_1,p_2^{\rm cr})$ instead of $G_0(p_1,p_2)$ alone.
Unlike the Salpeter version,
now all four possible projection operators $\Lambda_{12}^{(++)}$,
$\Lambda_{12}^{(+-)}$, $\Lambda_{12}^{(-+)}$, $\Lambda_{12}^{(--)}$,
enter the expression of
$\tilde G_0^{\rm MW}(M_B;{\bf p})$, Eq.~(\ref{G0-MW-1}).
For this reason, the inverse operator for the free Green function
in the 3D space exists.
further, in analogy with Eq.~(\ref{BSwf-1}),
we can define the inverse of the full Green function in 3D space according to
\eq\label{MW-full-G}
\bigl[\tilde G_0^{\rm MW}\bigr]^{-1}(M_B;{\bf p},{\bf p}')
-\bigl[\tilde G^{\rm MW}\bigr]^{-1}(M_B;{\bf p},{\bf p}')
=V({\bf p},{\bf p}')\, ,
\en
where
\eq\label{G0-MW-3}
\tilde G_0^{\rm MW}(M_B;{\bf p},{\bf p}')=(2\pi)^3\delta^3({\bf p}-{\bf p}')\,
\tilde G_0^{\rm MW}(M_B;{\bf p})\, .
\en
The MW equation for the bound-state vector
$|\tilde\Phi_{M_B}\rangle$ is given by
\eq\label{MW-wf-1}
\bigl[\tilde G^{\rm MW}\bigr]^{-1}|\tilde\Phi_{M_B}\rangle=0\quad
\Rightarrow\quad |\tilde\Phi_{M_B}\rangle
=\tilde G_0^{\rm MW}V|\tilde\Phi_{M_B}\rangle\, .
\en
Note that we can rewrite the inverse of the
free Green function in the MW equation as
\eq\label{MW-Green}
\bigl[\tilde G_0^{\rm MW}\bigr]^{-1}=\gamma_0^{(1)}\otimes\gamma_0^{(2)}\,
\biggl[ (E_1-h_1)\otimes\frac{h_2}{w_2}+(E_2-h_2)\otimes\frac{h_1}{w_1}\biggr]\, .
\en

With the use of this identity, one can rewrite the MW equation as
\eq\label{MW-wf-new}
\biggl[ (E_1-h_1)\otimes\frac{h_2}{w_2}+(E_2-h_2)\otimes\frac{h_1}{w_1}\biggr]\,
|\tilde\Phi_{M_B}\rangle=\gamma_0^{(1)}\otimes\gamma_0^{(2)}\, V\,
|\tilde\Phi_{M_B}\rangle\, .
\en

Let us now consider the limit of this equation when $m_1\rightarrow\infty$
(see Eq.~(\ref{limit})).
In this limit, according to Eq.~(\ref{WG-1}), $E_1\rightarrow m_1$,
$E_2\rightarrow M_B-m_1$, and the equation~(\ref{MW-wf-new}) simplifies to
the Dirac equation
\eq\label{MW-Dirac}
\bigl[ E_2-h_2\bigr]\,|\tilde\Phi_{M_B}\rangle=\gamma_0^{(2)}V
|\tilde\Phi_{M_B}\rangle\, .
\en
Consequently, the MW equation has the correct one-body limit.

\subsection{The Cooper-Jennings equation [9]}

The parameters $\alpha(s)$ and $\beta(s)$ in the Cooper-Jennings (CJ) version
are chosen as
\eq\label{CJ-ab}
&&\alpha(s)=\frac{\alpha_1(s)}{\alpha_1(s)+\alpha_2(s)}\, ,\quad\quad
\beta(s)=\frac{\alpha_2(s)}{\alpha_1(s)+\alpha_2(s)}\, ,
\nonumber\\[2mm]
&&\alpha_1(s)=\frac{s+m_1^2-m_2^2}{2\sqrt{s}}\, ,\quad\quad
\alpha_2(s)=\frac{s-m_1^2+m_2^2}{2\sqrt{s}}\, .
\en
The free Green function for the CJ equation is given by
\eq\label{CJ-G0}
G_0^{\rm CJ}(P;p)=-(\not\! p_1+m_1)\otimes (\not\! p_2+m_2)
g_0^{\rm CJ}(P;p)\, ,
\en
where $g_0^{\rm CJ}(P;p)$ is constrained by the elastic unitarity and
can be written in the following form
\eq\label{CJ-unitarity}
&&g_0^{\rm CJ}(P;p)=2\pi i\int_{(m_1+m_2)^2}^\infty
\frac{ds'f(s,s')}{s'-s-i0}\,\,
\delta^+\bigl[(\alpha(s')P'+p)^2-m_1^2\bigr]\,
\delta^+\bigl[(\beta(s')P'-p)^2-m_2^2\bigr]\, .
\nonumber\\
&&
\en
Here $\delta^+(x^2-a^2)=(2a)^{-1}\delta(x-a)$, $P'=\sqrt{s'/s}P$,
and the function $f(s,s')$ satisfies the condition $f(s,s)=1$.

After integration, the expression (\ref{CJ-unitarity}) yields
\eq\label{CJ-unitarity-1}
&&g_0^{\rm CJ}(P;p)=-2\pi i
\frac{\delta (2Pp)}{s-s_p}\,\,
\sqrt{s s_p}\,\, \frac{f(s,s_p)}{s-s_p} \alpha_1(s_p)\,\alpha_2(s_p)\, ,
\en
where $s_p=(\sqrt{m_1^2-p^2}+\sqrt{m_2^2-p^2})^2$.

Choosing the function $f(s,s_p)$ in the form
\eq\label{f-s-sp}
f(s,s_p)=\frac{4s\,\alpha_1(s_p)\,\alpha_2(s_p)}{ss_p-(m_1^2-m_2^2)^2}\, 
\en
we arrive at the following expression for $g_0^{\rm CJ}(P;p)$
\eq\label{CJ-g_0-int}
g_0^{\rm CJ}(P;p)=-2\pi i\,\,\frac{2s}{(s-(w_1+w_2)^2+i0)(s-(w_1-w_2)^2)}\,\,
\frac{2\sqrt{s}\,\delta(2P\cdot p)}{w_1+w_2}\, .
\en
Note, that in the c.m. frame, $2\sqrt{s}\,\delta(2P\cdot p)=\delta(p_0)$.
Because of the presence of the $\delta$ function, one can rewrite the free
Green function from~(\ref{CJ-G0}) in the following form (again, in the c.m. frame)
\eq\label{CJ-G0-1}
G_0^{\rm CJ}(M_B;p)=-(\not\! \tilde{p}_1+m_1)\otimes (\not\! \tilde{p}_2+m_2)
g_0^{\rm CJ}(M_B;p)\, ,\quad
\tilde p_1=(E_1,{\bf p})\, ,\,\,
\tilde p_2=(E_2,-{\bf p})\, ,
\en
where $E_1$ and $E_2$ are given by Eq.~(\ref{WG-1}).
The free Green function for the CJ equation in 3D space is related to 4D
Green function according to
\eq\label{CJ-G0-3-4}
G_0^{\rm CJ}(M_B;p)=2\pi i\,\delta(p_0)\,\tilde G_0^{\rm CJ}(M_B;{\bf p})\, ,
\en
where $\tilde G_0^{\rm CJ}(M_B;{\bf p})$ is given by
\eq\label{CJ-tilde-G0}
\tilde G_0^{\rm CJ}(M_B;{\bf p})
&=&\frac{1}{2(w_1+w_2)}\,\frac{(\not\! \tilde{p}_1+m_1)\otimes (\not\! \tilde{p}_2+m_2)}{E_1^2-w_1^2}
=\frac{1}{2(w_1+w_2)}\,\frac{(\not\! \tilde{p}_1+m_1)\otimes (\not\! \tilde{p}_2+m_2)}{E_2^2-w_2^2}
\nonumber\\[2mm]
&=&\frac{1}{2(w_1+w_2)}\,\frac{\not\! \tilde{p}_1+m_1}{\not\! \tilde{p}_2-m_2}
=\frac{1}{2(w_1+w_2)}\,\frac{\not\! \tilde{p}_2+m_2}{\not\! \tilde{p}_1-m_1}\, .
\en
In the limit, when one of the masses tends to infinity,
\eq\label{CJ-infty}
\frac{\not\! \tilde{p}_i+m_i}{2(w_1+w_2)}\rightarrow 1\quad\quad
{\rm at}\quad m_i\rightarrow\infty,\quad i=1,2\, .
\en
Consequently, the CJ equation has the correct one-body limit.

Note that, using the properties of the projection operators,
the free Green function in the 3D space can be rewritten in the following form
\eq\label{CJ-G0-3D}
\tilde G_0^{\rm CJ}(M_B;{\bf p})&=&\frac{1}{2(w_1+w_2)a}\,\bigl[
 (w_1+E_1)(w_2+E_2)\Lambda_{12}^{(++)}({\bf p},-{\bf p})
\nonumber\\[2mm]
&-&(w_1+E_1)(w_2-E_2)\Lambda_{12}^{(+-)}({\bf p},-{\bf p})
-(w_1-E_1)(w_2+E_2)\Lambda_{12}^{(-+)}({\bf p},-{\bf p})
\nonumber\\[2mm]
&+&(w_1-E_1)(w_2-E_2)\Lambda_{12}^{(--)}({\bf p},-{\bf p})\bigr]\,
\gamma_0^{(1)}\otimes\gamma_0^{(2)}\, ,
\en
where
\eq\label{CJ-par}
a=E_1^2-w_1^2=E_2^2-w_2^2=\frac{1}{4}\,\bigl[ M_B^2+b_0^2-2(w_1^2+w_2^2)\bigr]\, ,\quad
b_0=E_1-E_2=\frac{m_1^2-m_2^2}{M_B}\, .
\en

\subsection{The Maung-Norbury-Kahana equation [10,11]}

The free Green function for the Maung-Norbury-Kahana (MNK) equation is again given
by Eq.~(\ref{CJ-G0}), but with
\eq\label{MNK-g_0}
g_0^{\rm MNK}(P;p)=-2\pi i\,\frac{\delta^+\biggl\{
 \bigl[(\alpha(s)P+p)^2-m_1^2\bigr]\frac{1+y}{2}
-\bigl[(\beta(s)P-p)^2-m_2^2\bigr]\frac{1-y}{2}\biggr\} }{
 \bigl[(\alpha(s)P+p)^2-m_1^2\bigr]+\bigl[(\beta(s)P-p)^2-m_2^2\bigr]+i0}\, ,
\en
with $y=(m_1-m_2)/(m_1+m_2)$. This Green functions, of course, satisfies the unitariry
condition in the elastic channel. In addition, it has the property that the particles
1 and 2 in the intermediate states are now allowed to go off mass shell
inverse proportionally to their masses - so that, if one of the particles
becomes infinitely massive, it is automatically kept on its mass shell.

After some transformations, the Green function from Eq.~(\ref{MNK-g_0})
in the c.m. frame can be rewritten as
\eq\label{MNK-g_0-1}
&&g_0^{\rm MNK}(M_B;p)=-2\pi i\, \frac{\delta(p_0-p_0^+)}{2R(p_0^+p_0^-+a)}\, ,\quad
p_0^+=\frac{R-b}{2y}\, ,\quad
p_0^-=p_0^++b_0\, ,
\nonumber\\[2mm]
&&R=\sqrt{b^2-4y^2a}\, ,\quad
b=M_B+b_0y\, ,
\en
and $b_0$ is given by Eq.~(\ref{CJ-par}). By using the above expression,
we obtain
\eq\label{MNK-G0-3D}
\tilde G_0^{\rm MNK}(M_B,{\bf p})=
\frac{(\not\! \tilde p_1^++m_1)(\not\! \tilde p_2^++m_2)}{2R(p_0^+p_0^-+a)}
\, , \quad
\tilde p_1^+=(E_1+p_0^+,{\bf p}),\quad
\tilde p_2^+=(E_2-p_0^+,-{\bf p})\, .
\en

The relation between the 4D and 3D free Green functions in the MNK version
is given by
\eq\label{MNK-3D-4D}
G_0^{\rm MNK}(M_B;p)=2\pi i\delta(p_0-p_0^+)\,
\tilde G_0^{\rm MNK}(M_B,{\bf p})\, .
\en
Using the properties of the projection operators, the free Green function of 
the MNK equation can be recast in the following form
\eq\label{MNK-proj}
\tilde G_0^{\rm MNK}(M_B,{\bf p})&=&\frac{1}{2R(p_0^+p_0^-+a)}\,\,
\biggl\{\bigl[
 (w_1+E_1)(w_2+E_2)\Lambda_{12}^{(++)}({\bf p},-{\bf p})
\nonumber\\[2mm]
&-&(w_1+E_1)(w_2-E_2)\Lambda_{12}^{(+-)}({\bf p},-{\bf p})
-(w_1-E_1)(w_2+E_2)\Lambda_{12}^{(-+)}({\bf p},-{\bf p})
\nonumber\\[2mm]
&+&(w_1-E_1)(w_2-E_2)\Lambda_{12}^{(--)}({\bf p},-{\bf p})
\bigr]
\nonumber\\[2mm]
&-&\bigl[ p_0^+p_0^-
+(w_1-w_2)p_0^+(\Lambda_{12}^{(++)}({\bf p},-{\bf p})
               -\Lambda_{12}^{(--)}({\bf p},-{\bf p}))
\nonumber\\[2mm]
&+&(w_1+w_2)p_0^+(\Lambda_{12}^{(+-)}({\bf p},-{\bf p})
               -\Lambda_{12}^{(-+)}({\bf p},-{\bf p}))
\bigr]\biggr\}\gamma_0^{(1)}\otimes\gamma_0^{(2)}\, .
\en

In the limit when $m_1\rightarrow\infty$, the function $g_0^{\rm MNK}(P;p)$
from Eq.~(\ref{MNK-g_0}) is reduced to
\eq\label{MNK-g_0-infty}
g_0^{\rm MNK}(P;p)\biggr|_{m_1\rightarrow\infty}\rightarrow
\frac{-2\pi i}{p_2^2-m_2^2}\,\,\frac{\delta(p_0)}{2m_1}\, .
\en
From Eq.~(\ref{CJ-G0}) we can evaluate $G_0^{\rm MNK}(P;p)$ in this limit:
\eq\label{MNK-G0-infty}
G_0^{\rm MNK}(P;p)\biggr|_{m_1\rightarrow\infty}\rightarrow
2\pi i\,\,\frac{\delta(p_0)}{2m_1}\,\,
\frac{(\not\! {\tilde p_1}+m_1)\otimes (\not\! {\tilde p_2}+m_2)}{\tilde p_2^2-m_2^2}
\, ,
\en
where $\tilde p_1$ and $\tilde p_2$ are defined by Eq.~(\ref{CJ-G0-1}).
Integrating this relation over $p_0$, for the 3D free Green function 
in the c.m. frame we obtain
\eq\label{MNK-3D-G0}
\tilde G_0^{\rm MNK}(M_B;{\bf p})\biggr|_{m_1\rightarrow\infty}\rightarrow
\frac{\not\!{\tilde p_1}+m_1}{2m_1}\otimes
\frac{\not\!{\tilde p_2}+m_2}{\tilde p_2^2-m_2^2}\, .
\en
Since the factor $(\not\!{\tilde p_1}+m_1)/(2m_1)$ tends to unity in the limit
$m_1\rightarrow\infty$, one concludes that the MNK equation has the correct 
one-body limit.

\subsection{The normalization condition for the wave function
in MW, CJ and MNK versions}

The 3D free Green function in either of MW, CJ, or MNK versions, in the
c.m. frame can be rewritten in terms of the projection operators:
\eq\label{tG0-3}
\tilde G_0(M_B,{\bf p})=\sum_{\alpha_1,\alpha_2=\pm}
\frac{D^{(\alpha_1\alpha_2)}(M_B;p)}{d(M_B;p)}\,\,
\Lambda_{12}^{(\alpha_1\alpha_2)}({\bf p},-{\bf p})\,\,
\gamma_0^{(1)}\otimes\gamma_0^{(2)}\, ,\quad\quad
p=|{\bf p}|\, ,
\en
where
\eq\label{Dd}
{\rm MW} &\quad : \quad&
D^{(\alpha_1\alpha_2)}=\frac{(-)^{\alpha_1+\alpha_2}}
{(w_1+w_2)-(\alpha_1 E_1+\alpha_2 E_2)}\, ,\quad\quad d=1\, ,
\nonumber\\[2mm]
{\rm CJ} &\quad : \quad&
D^{(\alpha_1\alpha_2)}=(E_1+\alpha_1 w_1)(E_2+\alpha_2 w_2)\, ,\quad\quad
d=2(w_1+w_2) a\, ,
\nonumber\\[2mm]
{\rm MNK} &\quad : \quad&
D^{(\alpha_1\alpha_2)}=(E_1+\alpha_1 w_1)(E_2+\alpha_2 w_2)
\nonumber\\[2mm]
&&\quad\quad\quad\,\,
-\frac{R-b}{2y}\,\biggl(\frac{R-b}{2y}+(E_1+\alpha_1 w_1)
-(E_2+\alpha_2 w_2)\biggr)\, ,
\nonumber\\[2mm]
&&
d=2RB\, ,\quad
B=\frac{R-b}{2y}\biggl(\frac{R-b}{2y}+b_0\biggr)+a\, ,
\en
with $E_1,~E_2,~a,~b_0,~R,~b,~y$ defined above.

The equation for the bound-state wave function frequency components~(\ref{frequency}) 
can be directly obtained from Eq.~(\ref{Sal-1}) by substituting the
above expression for the free 3D Green function and using the properties of
the projection operators
\eq\label{frequency-wf-3}
\bigl[ M_B-(\alpha_1 w_1+\alpha_2 w_2)\bigr]
\tilde\Phi^{(\alpha_1\alpha_2)}_{M_B}({\bf p})&=&
A^{(\alpha_1\alpha_2)}(M_B;p)\,
\Lambda_{12}^{(\alpha_1\alpha_2)}({\bf p},-{\bf p})\,
\gamma_0^{(1)}\otimes\gamma_0^{(2)}\times
\nonumber\\[2mm]
&\times&\sum_{\alpha_1'\alpha_2'}\int\frac{d^3{\bf p}'}{(2\pi)^3}\,
V({\bf p},{\bf p}')\,\tilde\Phi^{(\alpha_1'\alpha_2')}_{M_B}({\bf p}')\, ,
\en 
where, for the different versions
\eq\label{A-3}
{\rm MW} &\quad : \quad&
A^{(\pm\pm)}=1\, ,\quad A^{(\pm\mp)}=\frac{M_B}{w_1+w_2}\, ,
\nonumber\\[2mm]
{\rm CJ} &\quad : \quad&
A^{(\alpha_1\alpha_2)}=\frac{M_B+(\alpha_1 w_1+\alpha_2 w_2)}{2(w_1+w_2)}\, ,
\nonumber\\[2mm]
{\rm MNK} &\quad : \quad&
A^{(\alpha_1\alpha_2)}=\frac{1}{2RB}\,\biggl\{
a\bigl[ M_B+(\alpha_1 w_1+\alpha_2 w_2)\bigr]
-\bigl[ M_B-(\alpha_1 w_1+\alpha_2 w_2)\bigr] \frac{R-b}{2y}\times
\nonumber\\[2mm]
&&\quad\quad\quad\,\,
\times\biggl(\frac{R-b}{2y}+(E_1+\alpha_1 w_1)
-(E_2+\alpha_2 w_2)\biggr)\biggr\}\, .
\en
Thus, the MW, CJ and MNK equations couple all four frequency components of the
wave function: $\tilde\Phi^{(++)}_{M_B}$, $\tilde\Phi^{(+-)}_{M_B}$,
$\tilde\Phi^{(-+)}_{M_B}$ and $\tilde\Phi^{(--)}_{M_B}$. One can formally 
extend these notations for the Salpeter (SAL) and Gross (GR) versions,
defining
\eq\label{A-4_5}
{\rm SAL} &\quad : \quad&
A^{(\pm\pm)}=\pm 1\, ,\quad A^{(\pm\mp)}=0\, ,
\nonumber\\[2mm]
{\rm GR} &\quad : \quad&
A^{(+\pm)}=\pm 1\, ,\quad A^{(-\mp)}=0\, .
\en
It is immediately seen that Salpeter and Gross equations couple only two 
frequency components of the wave function, other two being equal to $0$.

For the derivation of the wave function normalization condition 
in MW, CJ and MNK versions, let us consider the full 3D Green function
that obeys the equation
\eq\label{tG-3}
\tilde G^i=\tilde G_0^i+\tilde G_0^i V\tilde G^i
=\tilde G_0^i+\tilde G^i V\tilde G_0^i\, ,\quad\quad
i={\rm MW,~CJ,~MNK}\, .
\en

In analogy with Eq.~(\ref{boundstatepole}), this Green function develops a 
bound-state pole(s)
\eq\label{3D-boundstatepole}
\langle{\bf p}|\tilde G(P)|{\bf p}'\rangle=
\sum_B\frac{\langle{\bf p}|\tilde\Phi_{{\bf P}_B}\rangle
\langle\tilde{\bar\Phi}_{{\bf P}_B}|{\bf p}'\rangle}
{P^2-M_B^2}+\langle{\bf p}|\tilde R(P)|{\bf p}'\rangle\, .
\en

This leads to the normalization condition in MW, CJ and MNK versions
in analogy with Eq.~(\ref{normalization})
\eq\label{3D-normalization}
\int\frac{d^3{\bf p}}{(2\pi)^3}\,\frac{d^3{\bf p}'}{(2\pi)^3}\,
\tilde{\bar\Phi}^i_{M_B}({\bf p})\,
\biggl\{\frac{\partial}{\partial M_B}\biggl(
(\tilde G_0^i(M_B;{\bf p},{\bf p}'))^{-1}-V({\bf p},{\bf p}')\biggr)
\biggr\}\tilde\Phi^i_{M_B}({\bf p}')=2M_B\, .
\en

Now using the relation
\eq\label{G-3D-3}
(\tilde G_0^i(M_B;{\bf p},{\bf p}'))^{-1}=
(2\pi)^3\delta^3({\bf p}-{\bf p}')\,(\tilde G_0^i(M_B;{\bf p}))^{-1}\, ,
\en
and the fact that $V({\bf p},{\bf p}')$ does not depend on $M_B$,
the normalization condition can be rewritten as
\eq\label{3D-normalization-1}
\int\frac{d^3{\bf p}}{(2\pi)^3}\,\tilde{\bar\Phi}^i_{M_B}({\bf p})\,
\biggl\{\frac{\partial}{\partial M_B}\,(\tilde G_0^i(M_B;{\bf p}))^{-1}
\biggr\}\,\tilde\Phi^i_{M_B}({\bf p})=2M_B\, .
\en

If now one substitutes here the expression for the free Green function
given by Eq.~(\ref{tG0-3}), one obtains (below, we drop the superscript
``i'' labeling various versions)
\eq\label{3D-normalization-2}
\sum_{\alpha_1\alpha_2}\int\frac{d^3{\bf p}}{(2\pi)^3}\,
\tilde{\bar\Phi}^{(\alpha_1\alpha_2)}_{M_B}({\bf p})\,
f_{12}^{(\alpha_1\alpha_2)}(M_B;p)\,
\tilde \Phi^{(\alpha_1\alpha_2)}_{M_B}({\bf p})=2M_B\, ,
\en
where
\eq\label{f_12}
f_{12}^{(\alpha_1\alpha_2)}(M_B;p)=\frac{\partial}{\partial M_B}
\biggl(\frac{d(M_B;p)}{D^{\alpha_1\alpha_2}(M_B;p)}\biggr)\, .
\en

By using the explicit expressions for $D^{(\alpha_1\alpha_2)}$ and $d$
given by Eq.~(\ref{Dd}), we obtain
\eq\label{f_12-Dd}
{\rm MW} &\quad : \quad&
f_{12}^{(\alpha_1\alpha_2)}=\frac{\alpha_1 E_1+\alpha_2 E_2}{M_B}\, ,
\nonumber\\[2mm]
{\rm CJ} &\quad : \quad&
f_{12}^{(\alpha_1\alpha_2)}=
\frac{2(w_1+w_2)}{M_B}\,\frac{\alpha_1 w_1 E_1+\alpha_2 w_2 E_2}
{(E_1+\alpha_1 w_1)(E_2+\alpha_2 w_2)}\, ,
\nonumber\\[2mm]
{\rm MNK} &\quad : \quad&
f_{12}^{(\alpha_1\alpha_2)}=\frac{2}{D^{(\alpha_1\alpha_2)}}\,
\biggl\{\biggl[\frac{M_B B}{R}(1-y^2)+\frac{M_B^2}{2}
-2\biggl(\frac{R-M_B}{2y}\biggr)^2\biggr]
\nonumber\\[2mm]
&-&\frac{B}{D^{(\alpha_1\alpha_2)}}\,\biggl[\biggl(M_B
+\frac{\alpha_1 w_1+\alpha_2 w_2}{2}\biggr) R -\frac{M_B^2}{2}
+2\biggl(\frac{R-M_B}{2y}\biggr)^2
\nonumber\\[2mm]
&+&(\alpha_1 w_1-\alpha_2 w_2)
\biggl(\frac{R-M_B}{2y}+\frac{M_B y}{2}\biggr)\biggr]\biggr\}\, .
\en

Let us note that the normalization condition~(\ref{3D-normalization-2})
is valid for the Salpeter and Gross versions as well, provided we choose
\eq\label{f_12-SG}
{\rm SAL} &\quad : \quad&
f_{12}^{(\alpha_1\alpha_2)}=\frac{\alpha_1+\alpha_2}{2}\, ,
\nonumber\\[2mm]
{\rm GR} &\quad : \quad&
f_{12}^{(+\pm)}=1\, ,\quad f_{12}^{(-\pm)}=0\, .
\en

Let us emphasize that the Salpeter, MW, CJ and MNK equations can be used for 
the bound systems with the equal masses of the constituents, whereas the
Gross equation can not - the particle ``1'' (spectator) should be heavier
than the particle ``2''. This is due to the 
approximation~(\ref{approximation-Gross}) that was done in the free Green
function of the Gross equation. Further, it is directly seen from
Eqs.~(\ref{A-3}) and (\ref{A-4_5}), that the Salpeter and Gross equations 
are linear eigenvalue equations for determining $M_B$ (the functions
$A^{(\alpha_1\alpha_2)}$ do not depend on $M_B$), whereas MW, CJ and MNK
equations are not, and $M_B$ enters the right-hand side of these equations 
as well.

Let us now concentrate on the properties of the coefficient functions
in detail. In the case of the equal-mass constituents $m_1=m_2=m$ and
$w_1=w_2=w$, we obtain
\eq\label{A-equal}
{\rm MW} &\quad : \quad&
A^{(\pm\pm)}=1\, ,\quad A^{(\pm\mp)}=\frac{M_B}{2w}\, ,
\nonumber\\[2mm]
{\rm CJ} &\quad : \quad&
A^{(\pm\pm)}=\frac{M_B\pm 2w}{4w}\, ,\quad A^{(\pm\mp)}=\frac{M_B}{4w}\, ,
\nonumber\\[2mm]
{\rm MNK} &\quad : \quad&
A^{(\pm\pm)}=\frac{M_B\pm 2w}{2M_B}\, ,\quad A^{(\pm\mp)}=\frac{1}{2}\, .
\en

It is immediately seen that in the equal-mass case, the $M_B$ drops out
from the equations for mixed components $\tilde\Phi_{M_B}^{(\pm\mp)}$ in
the MW and CJ versions - that is, these components are redundant and 
can be eliminated in this case.

The functions $f_{12}^{(\alpha_1\alpha_2)}$ in the equal-mass case
are given by
\eq\label{f_12-equal}
{\rm MW} &\quad : \quad&
f_{12}^{(\pm\pm)}=\pm 1\, ,\quad f_{12}^{(\pm\mp)}=0\, ,
\nonumber\\[2mm]
{\rm CJ} &\quad : \quad&
f_{12}^{(\pm\pm)}=\pm \frac{16w^2}{(M_B\pm 2w)^2}\, 
,\quad f_{12}^{(\pm\mp)}=0\, ,
\nonumber\\[2mm]
{\rm MNK} &\quad : \quad&
f_{12}^{(\pm\pm)}=\frac{2((M_B\pm 2w)^2-8w^2)}{(M_B\pm 2w)^2}\, 
,\quad f_{12}^{(\pm\mp)}=2\, .
\en

From these expressions, we immediately see that in the CJ and MNK versions 
the function $f_{12}^{(--)}$ has the second-order pole at
\eq\label{normalization-pole}
M_B-2w(p_s)=0\, ,\quad\quad p_s=\frac{1}{2}\,\sqrt{M_B^2-4m^2}\, .
\en

It can be shown that in the non-equal mass case in the CJ and MNK versions
the function $f_{12}^{(--)}$ has the second-order pole at
\eq\label{norm-pole-uneq}
E_i^2-w_i^2(p_s)=0\, ,\quad\quad
p_s=\frac{1}{2}\,\sqrt{M_B^2+\biggl(\frac{m_1^2-m_2^2}{M_B}\biggr)^2-
2(m_1+m_2)^2}\, ,
\en
while the other components  $f_{12}^{(++)}$, $f_{12}^{(+-)}$ 
and $f_{12}^{(-+)}$ do not have any poles.

\subsection{Logunov-Tavkhelidze quasipotential approach [12]}

There exists the theoretical possibility to construct the 3D analogue
of the BS equation without using the instantaneous approximation. To
this end, one may use the Logunov-Tavkhelidze quasipotential
approach formulated in Ref.~\cite{Logunov} for the case of two
spinless particles, and generalized in Ref.~\cite{Khelashvili} to the
case of two fermions.

We introduce the following definition. For any operator $A(P)$ in the
momentum space,
\eq\label{tilde}
\langle{\bf p}|\tilde A(P)|{\bf p}'\rangle=
\int\frac{dp_0}{2\pi}\,\frac{dp_0'}{2\pi}\,\langle p|A(P)|p'\rangle\, .
\en 

Then, from Eq.~(\ref{BSG}) one obtains
\eq\label{G-tilde}
\tilde G=\tilde G_0+\widetilde{G_0 K G}\, ,
\en
where $\tilde G_0$ is given by Eq.~(\ref{equal-time-Green}).

Due to the fact that the operator $\Pi$ defined by Eq.~(\ref{Pi}) can
not be inverted, the inverse operator of $\tilde G_0$ does not exist
as well. As a result, one can not define the interaction potential by
the formula analogous to Eq.~(\ref{MW-full-G}). In order to overcome
this problem, it is convenient to introduce the Green function
$\underline{\tilde G_0}$ defined by

\eq\label{underline-G0}
\underline{\tilde G_0}(P;{\bf p})&=&
i\bigl[ P_0-h_1({\bf p}_1)-h_2({\bf p}_2\bigr]^{-1}
\gamma_0^{(1)}\otimes\gamma_0^{(2)}\, ,
\nonumber\\[2mm]
\tilde G_0(P;{\bf p},{\bf p}')&=&(2\pi)^3\delta^3({\bf p}-{\bf p}')\,
\underline{\tilde G_0}(P;{\bf p})\,
\gamma_0^{(1)}\otimes\gamma_0^{(2)}\,\hat \Pi\,
\en
where $\hat\Pi=\Pi\gamma_0^{(1)}\otimes\gamma_0^{(2)}$.

Now, the inverse of the operator 
$\underline{\tilde G_0}(P;{\bf p},{\bf p}')
=(2\pi)^3\delta^3({\bf p}-{\bf p}')\,\underline{\tilde G_0}(P;{\bf p})$
exists, and one may define
\eq\label{underline-G}
\underline{\tilde G}=\underline{\tilde G_0}+\widetilde{G_0KG}\, ,
\en
from which follows that
\eq\label{G-underline-G}
\underline{\tilde G}=\tilde G+\underline{\tilde G_0}\,
(1-\gamma_0^{(1)}\otimes\gamma_0^{(2)}\hat\Pi)\, .
\en

It is clear, that near the bound-state pole the Green functions 
$\tilde G$ and $\underline{\tilde G}$ differ only by the regular term,
since $\underline{\tilde G_0}$ is regular in the vicinity of the
pole. Consequently, in order to derive the bound-state equation, one
may use $\underline{\tilde G}$ instead of $\tilde G$, and define the
interaction potential according to
\eq\label{underline-V}
\bigl[\,\underline{\tilde G_0}\,\bigr]^{-1}-
\bigl[\,\underline{\tilde G}\,\bigr]^{-1}=\underline{\tilde V}\, ,
\en
from which it follows that
\eq\label{eq-underline}
\underline{\tilde G}=\underline{\tilde G_0}+
\underline{\tilde G_0}\, \underline{\tilde V}\, \underline{\tilde G}\, ,
\en
and
\eq\label{exp-underline-V}
\underline{\tilde V}=
\bigl[\,\underline{\tilde G_0}\,\bigr]^{-1}\,\widetilde{G_0 K G}\,
\bigl[\,\underline{\tilde G_0}\,\bigr]^{-1}\, .
\en 
For any given kernel $K(P;p,p')$ the interaction potential can be
constructed by using Eq.~(\ref{exp-underline-V}). The equation for the
bound-state wave function in the c.m. frame can be obtained directly from
Eq.~(\ref{underline-V})
\eq\label{underline-wf-1}
\underline{\tilde G}^{-1}(M_B)\, |\tilde\Phi_{M_B}\rangle=0\, ,\quad\quad
|\tilde\Phi_{M_B}\rangle=\underline{\tilde G}\,\underline{\tilde V}\,
|\tilde\Phi_{M_B}\rangle\, .
\en
Defining the quasipotential as
\eq\label{quasi}
V_q(M_B;{\bf p},{\bf p}')=i\gamma_0^{(1)}\otimes\gamma_0^{(2)}\,
\underline{\tilde V}(M_B;{\bf p},{\bf p}')\, ,
\en
we obtain
\eq\label{underline-wf-2}
\biggl[ M_B-h_1({\bf p})-h_2(-{\bf p})\biggr]\,
\tilde\Phi_{M_B}({\bf p})=
\int\frac{d^3{\bf p}'}{(2\pi)^3}\,\,V_q(M_B;{\bf p},{\bf p}')\,\,
\tilde\Phi_{M_B}({\bf p}')\, .
\en

Note that in the instantaneous approximation~(\ref{static}),
the interaction potential reduces to
\eq\label{V-st}
\underline{\tilde V}(M_B;{\bf p},{\bf p}')=
\gamma_0^{(1)}\otimes\gamma_0^{(2)}\,\Pi({\bf p})\,
\gamma_0^{(1)}\otimes\gamma_0^{(2)}\, K_{st}({\bf p},{\bf p}')\, .
\en
As a result, the quasipotential equation reduces to the Salpeter
equation.

The first-order quasipotential is defined by Eqs.~(\ref{underline-V})
and (\ref{quasi}), if in the former the full Green function $G$ is
substituted by the free Green function $G_0$
\eq\label{quasi-1}
V_q^{(1)}(M_B;{\bf p},{\bf p}')=i\gamma_0^{(1)}\otimes\gamma_0^{(2)}\,\,
\langle {\bf p}|\bigl[\,\underline{\tilde G_0}\,\bigr]^{-1}\,
\widetilde{G_0 K G}\,\bigl[\,\underline{\tilde G_0}\,\bigr]^{-1}|
{\bf p}'\rangle\, 
\en
from which, in the static approximation one obtains
\eq\label{quasi-1-st}
V_q^{(1)}(M_B;{\bf p},{\bf p}')=\Pi({\bf p})\,
\gamma_0^{(1)}\otimes\gamma_0^{(2)}\,\, iK_{st}({\bf p},{\bf p}')\, .
\en

It is seen that, unlike the full quasipotential equation, the
first-order equation does not reduce to the Salpeter equation in the
static limit. Only when one may neglect the negative-frequency
component of the bound-state wave function, the first-order equation
again reduces to the Salpeter equation in the static limit.
Here we note, that the first-order
quasipotential equation was used in Ref.~\cite{KR} in order to
evaluate the dynamical retardation effect in the $q\bar q$ bound
system mass spectrum (i.e. the effect that stems from the deviation of
the BS kernel from the static one).

In the rest of this subsection, we consider the normalization
condition for the quasipotential bound-state wave function.
Near the bound-state pole, the 3D Green function $\tilde G(P)$ develops a
pole~(\ref{3D-boundstatepole}). Using the fact that in the vicinity of
the bound-state pole the Green functions $\tilde G(P)$ and
$\underline{\tilde G(P)}$ coincide up to the regular term, it is
straightforward to obtain the normalization condition
\eq\label{norm-underline}
i\int\frac{d^3{\bf p}}{(2\pi)^3}\,\frac{d^3{\bf p}'}{(2\pi)^3}\,
\tilde{\bar\Phi}_{M_B}({\bf p})\,\biggl[\frac{\partial}{\partial M_B}
\biggl( (\underline{\tilde G_0}(M_B;{\bf p},{\bf p}'))^{-1}
-\underline{\tilde V}(M_B;{\bf p},{\bf p}')\biggr)
\tilde\Phi_{M_B}({\bf p}')=2M_B\, .
\en
From this equation, using the definition of the conjugate wave
function~(\ref{w.f.}) and Eqs.~(\ref{underline-G0}), (\ref{quasi}), we obtain
\eq\label{norm-quasi}
\int\frac{d^3{\bf p}}{(2\pi)^3}\,\frac{d^3{\bf p}'}{(2\pi)^3}\,
\tilde\Phi^\dagger_{M_B}({\bf p})\,\biggl[ {\bf 1} - 
\frac{\partial}{\partial M_B}\biggl( V_q(M_B;{\bf p},{\bf p}')\biggr)\biggr]
\tilde\Phi_{M_B}({\bf p}')=2M_B\, .
\en

As it is seen from Eq.~(\ref{norm-quasi}), in the static limit the 
above normalization condition
reduces to the normalization condition for the Salpeter wave function
only if one neglects the contribution from the negative-energy
component of the wave function.

\section{Meson spectroscopy}
\setcounter{equation}{0}

\subsection{Partial-wave decomposition}

The properties of the $q\bar q$ bound systems in the 3D formalism
obtained from the BS equation in the static approximation, were
studied in Refs.~\cite{PLB,Long,AChK,ChK-Czech,ChK,AChK-Yad,Lagae,AChK-FBS,Vary,Tjon,Resag-1,Resag-2,AChKR,Parramore,Resag-3,Resag-4,Zoller,Olsson,Klempt,Parramore-1,Munz,BKR},
without making any additional assumptions.
Note that these 3D equations can be written either, as in
Eq.~(\ref{Sal-2}), for the 2-fermion bound-state wave function~\cite{PLB,Long,AChK,ChK-Czech,ChK,AChK-Yad,AChK-FBS,Vary,Tjon,AChKR,Parramore,Parramore-1,BKR},
or for the fermion-antifermion bound-state wave function~\cite{Lagae,Resag-1,Resag-2,Resag-3,Resag-4,Zoller,Olsson,Klempt,Munz}
(the latter is obtained from Eq.~(\ref{BS-C}) in the static approximation.).
Further, one may write down these equations it terms of either the
frequency components of the 3D wave functions 
$\tilde\Phi^{(\pm\pm)}_{M_B}({\bf p})$ and
 $\tilde\Psi^{(\pm\pm)}_{M_B}({\bf p})$ 
(the latter denotes the frequency components of the
fermion-antifermion wave function), or on terms of their linear
combinations $\tilde\Phi_{aa}({\bf p}),~\tilde\Phi_{bb}({\bf p})$, etc,
$\tilde\Psi_{aa}({\bf p}),~\tilde\Psi_{bb}({\bf p})$, etc, see
Eqs.~(\ref{wf-2X2}), (\ref{Psi-C}).
Below, we shall use the form of the 3D equations given
by~(\ref{frequency-wf-3})-(\ref{A-4_5}), with the normalization
condition given by~(\ref{3D-normalization-2})-(\ref{f_12-SG})~\cite{PLB,AChKR,BKR}.

In order to rewrite the equations explicitly in either of the forms
above, one has to specify the explicit spin structure of the
interaction potential. This potential consists of several parts.
First, there is the one-gluon (OG) exchange piece dominating at short
distances. In the Feynman gauge, the spin structure of this piece
is given by $\gamma_\mu^{(1)}\otimes\gamma^{(2)\mu}=
\gamma_0^{(1)}\otimes\gamma_0^{(2)}
-\mathbold{\gamma}^{(1)}\otimes\mathbold{\gamma}^{(2)}$.
In accordance with the static approximation, however, we neglect the
second term in this expression~\cite{Itzykson}. In addition, there is the
confinement (C) piece in the potential that dominates at large
distances and leads to the formation of the $q\bar q$ bound states. The
spin structure of this piece is not known {\it a priori}. We choose
it to be the mixture of a scalar and a zeroth component of a
vector. Further, sometimes an additional ``instanton-induced'' piece 
corresponding to the t'Hooft interaction, is included
in the potential~\cite{Resag-2}. The spin structure of this term is
given by the equal mixture of scalar and pseudoscalar parts.
The rationale for including the latter piece is the following.
In the absence of the proper treatment of the Goldstone nature of
light pseudoscalar bosons that is due to the spontaneous breaking of
chiral symmetry in QCD, the t'Hooft interaction mimics this effect,
leading to the large mass splitting between the pseudoscalar and
vector mesons. Note that the chiral symmetry can be consistently
incorporated in the 3D framework (see, e.g.~\cite{Lagae}), albeit at
a cost of the more involved formalism. For example, in this case the
Hamiltonian of the free quark is replaced by
\eq\label{H-free}
h_i({\bf p}_i)=\mathbold{\alpha}^{(i)}{\bf p}_i+m_i\gamma_0^{(i)}\rightarrow 
B_i({\bf p}_i)\mathbold{\alpha}^{(i)}{\bf p}_i+A_i({\bf p}_i)\gamma_0^{(i)}\, ,
\en
where $A_i({\bf p}_i)$ and $B_i({\bf p}_i)$ are determined by solving
the gap equation for the quark propagator with the static potential.
Below, however, we do not consider this approach.

Thus, the spin structure of the static potential we shall be using, is
given by
\eq\label{V-static}
V&=&\gamma_0^{(1)}\otimes\gamma_0^{(2)}\, V_{\rm OG}(r)
+(x\gamma_0^{(1)}\otimes\gamma_0^{(2)}+(1-x) I^{(1)}\otimes I^{(2)})\,
V_{\rm C}(r)
\nonumber\\[2mm]
&+&(I^{(1)}\otimes I^{(2)}+\gamma_5^{(1)}\otimes\gamma_5^{(2)})\, 
V_{\rm T}(r)\, ,
\en
where the last term corresponds to the t'Hooft interaction, all
potentials are assumed to be local, and $0\leq x\leq 1$.
 
Let us now turn to the wave function. It is possible to ``solve'' the
constraints imposed on the frequency components, defining
\eq\label{Pauli}
\tilde\Phi_{M_B}^{(\alpha_1\alpha_2)}({\bf p})=
{\cal N}_{12}^{(\alpha_1\alpha_2)}(p)\,
\pmatrix{1\cr\frac{\alpha_1(\mathbold{\sigma}^{(1)}{\bf p})}{w_1+\alpha_1 m_1}}\otimes
\pmatrix{1\cr\frac{-\alpha_2(\mathbold{\sigma}^{(2)}{\bf p})}{w_2+\alpha_2 m_2}}\,\, 
\chi_{M_B}^{(\alpha_1\alpha_2)}({\bf p})=
\pmatrix{
\tilde\Phi_{aa}^{(\alpha_1\alpha_2)}({\bf p})\cr
\tilde\Phi_{ab}^{(\alpha_1\alpha_2)}({\bf p})\cr
\tilde\Phi_{ba}^{(\alpha_1\alpha_2)}({\bf p})\cr
\tilde\Phi_{bb}^{(\alpha_1\alpha_2)}({\bf p})}\, ,
\en
where
\eq\label{cal-N}
{\cal N}_{12}^{(\alpha_1\alpha_2)}(p)=
\sqrt{\frac{w_1+\alpha_1 m_1}{2w_1}}\,
\sqrt{\frac{w_2+\alpha_2 m_2}{2w_2}}=
{\cal N}_1^{(\alpha_1)}(p)\,{\cal N}_2^{(\alpha_2)}(p)\, ,
\en
and $\chi_{M_B}^{(\alpha_1\alpha_2)}({\bf p})$ is the unconstrained
Pauli $2\times 2$ spinor. For this spinor, the following system of
equations is obtained
\eq\label{eq-Pauli}
&&\bigl[ M_B-(\alpha_1 w_1+\alpha_2 w_2)\bigr]
\chi_{M_B}^{(\alpha_1\alpha_2)}({\bf p})=
A^{(\alpha_1\alpha_2)}(M_B;p)\sum_{\alpha_1'\alpha_2'}\int
\frac{d^3{\bf p}'}{(2\pi)^3}\, 
V_{\rm eff}^{(\alpha_1\alpha_2\alpha_1'\alpha_2')}({\bf p},{\bf p}')
\chi_{M_B}^{(\alpha_1'\alpha_2')}({\bf p}') ,
\nonumber\\
&&
\en
where
\eq\label{V-eff}
&&V_{\rm eff}^{(\alpha_1\alpha_2\alpha_1'\alpha_2')}({\bf p},{\bf p}')
={\cal N}_{12}^{(\alpha_1\alpha_2)}(p)\bigl(
 V_1({\bf p}-{\bf p}')
B_1^{(\alpha_1\alpha_2\alpha_1'\alpha_2')}({\bf p},{\bf p}')
+V_2(x;{\bf p}-{\bf p}')
B_2^{(\alpha_1\alpha_2\alpha_1'\alpha_2')}({\bf p},{\bf p}')
\nonumber\\[2mm]
&&+V_{\rm T}({\bf p}-{\bf p}')\,(
 B_1^{(\alpha_1\alpha_2\alpha_1'\alpha_2')}({\bf p},{\bf p}')
-B_2^{(\alpha_1\alpha_2\alpha_1'\alpha_2')}({\bf p},{\bf p}')
-B_3^{(\alpha_1\alpha_2\alpha_1'\alpha_2')}({\bf p},{\bf p}'))
\bigr)\, 
{\cal N}_{12}^{(\alpha_1'\alpha_2')}(p')
\en
and
\eq\label{B_i}
B_1^{(\alpha_1\alpha_2\alpha_1'\alpha_2')}({\bf p},{\bf p}')
&=&1+\frac{\alpha_1\alpha_2\alpha_1'\alpha_2'\,
(\mathbold{\sigma}^{(1)}{\bf p})(\mathbold{\sigma}^{(2)}{\bf p})
(\mathbold{\sigma}^{(1)}{\bf p}')(\mathbold{\sigma}^{(2)}{\bf p}')}
{(w_1+\alpha_1 m_1)(w_2+\alpha_2 m_2)
 (w_1'+\alpha_1' m_1)(w_2'+\alpha_2' m_2)}\, ,
\nonumber\\[2mm]
B_2^{(\alpha_1\alpha_2\alpha_1'\alpha_2')}({\bf p},{\bf p}')
&=&\frac{\alpha_1\alpha_1'\,(\mathbold{\sigma}^{(1)}{\bf p})(\mathbold{\sigma}^{(1)}{\bf p}')}
{(w_1+\alpha_1 m_1)(w_1'+\alpha_1' m_1)}
+\frac{\alpha_2\alpha_2'\,(\mathbold{\sigma}^{(2)}{\bf p})(\mathbold{\sigma}^{(2)}{\bf p}')}
{(w_2+\alpha_2 m_2)(w_2'+\alpha_2' m_2)}
\nonumber\\[2mm]
B_3^{(\alpha_1\alpha_2\alpha_1'\alpha_2')}({\bf p},{\bf p}')
&=&\frac{\alpha_1\alpha_2\,(\mathbold{\sigma}^{(1)}{\bf p})(\mathbold{\sigma}^{(2)}{\bf p})}
{(w_1+\alpha_1 m_1)(w_2+\alpha_2 m_2)}
+\frac{\alpha_1'\alpha_2'\,(\mathbold{\sigma}^{(1)}{\bf p}')(\mathbold{\sigma}^{(2)}{\bf p}')}
{(w_1'+\alpha_1' m_1)(w_2'+\alpha_2' m_2)}
\nonumber\\[2mm]
&-&\frac{\alpha_1\alpha_2'\,(\mathbold{\sigma}^{(1)}{\bf p})(\mathbold{\sigma}^{(2)}{\bf p}')}
{(w_1+\alpha_1 m_1)(w_2'+\alpha_2' m_2)}
-\frac{\alpha_1'\alpha_2\,(\mathbold{\sigma}^{(1)}{\bf p}')(\mathbold{\sigma}^{(2)}{\bf p})}
{(w_1'+\alpha_1' m_1)(w_2+\alpha_2 m_2)}
\en
\eq\label{V_i}
V_1=V_{\rm OG}+V_{\rm C}\, ,\quad\quad
V_2(x)=V_{\rm OG}+(2x-1)V_{\rm C}\, .
\en
The functions $V_{\rm OG}({\bf p}-{\bf p}')$, $V_{\rm C}({\bf p}-{\bf p}')$
and $V_{\rm T}({\bf p}-{\bf p}')$ are the Fourier-transform of the
local potentials $V_{\rm OG}(r)$, $V_{\rm C}(r)$ and $V_{\rm T }(r)$,
respectively.

The normalization condition for the Pauli spinors 
$\chi_{M_B}^{(\alpha_1\alpha_2)}({\bf p})$ follows
from~(\ref{3D-normalization-2})
\eq\label{normalization-chi}
\sum_{\alpha_1\alpha_2}\int\frac{d^3{\bf p}}{(2\pi)^3}\,
\chi_{M_B}^{\dagger(\alpha_1\alpha_2)}({\bf p})\,
f_{12}^{(\alpha_1\alpha_2)}(p)\,
\chi_{M_B}^{(\alpha_1\alpha_2)}({\bf p})=2M_B\, .
\en

The partial-wave expansion of the Pauli spinor 
$\chi_{M_B}^{(\alpha_1\alpha_2)}({\bf p})$ is given by
\eq\label{partial-wave}
\chi_{M_B}^{(\alpha_1\alpha_2)}({\bf p})=
\sum_{LSJM_J}\chi_{LSJM_J}^{(\alpha_1\alpha_2)}({\bf p})=
\sum_{LSJM_J}\langle {\bf n}|LSJM_J\rangle 
R^{(\alpha_1\alpha_2)}_{LSJM_J}(p)\, ,\quad
{\bf n}=\frac{{\bf p}}{p}\, ,
\en
where $R^{(\alpha_1\alpha_2)}_{LSJM_J}(p)$ denote the radial wave functions,
and $L,S,J,M_J$ stand for the total orbital angular momentum, total spin,
total angular momentum, and the projection of the total angular momentum of
the $q\bar q$ system, respectively.

The partial-wave expansion of the potentials reads as
\eq\label{PW-potentials}
V({\bf p}-{\bf p}')=(2\pi)^3\sum_{\bar L\bar S\bar J\bar M_{\bar J}}
\langle {\bf n}|\bar L\bar S\bar J\bar M_{\bar J}\rangle\, 
V_{\rm I}^{\bar L}(p,p')\,
\langle \bar L\bar S\bar J\bar M_{\bar J}|{\bf n}'\rangle\, ,\quad
{\rm I=OG,C,T}\, ,
\en
where
\eq\label{L-potentials}
V_{\rm I}^{\bar L}(p,p')=\frac{2}{\pi}\int_0^\infty
r^2 dr\,j_{\bar L}(pr)\, V_{\rm I}(r)\, j_{\bar L}(p'r)\, ,
\en
$j_{\bar L}$ being the spherical Bessel function.

Using the fact that for the spherical potentials
$V_{\rm I}({\bf p}-{\bf p}')=V_{\rm I}(|{\bf p}-{\bf p}'|)$, one may
write
\eq\label{L-potential-central}
V_{\rm I}^{\bar L}(p,p')=\frac{1}{4\pi^2}\int_{-1}^1 dz\,
P_{\bar L}(z)\, V_{\rm I}(\sqrt{p^2+{p'}^2-2pp'z})\, ,
\en
where $P_{\bar L}(z)$ denotes the Legendre polynomial. The above form is
convenient when the function $V_{\rm I}(p,p';z)$ can be written in the
analytic form.

In order to carry out the partial-wave expansion in the bound-state equation,
it is convenient to introduce the operators 
${\bf S}=\frac{1}{2}(\mathbold{\sigma}^{(1)}+\mathbold{\sigma}^{(2)})$, 
${\bf \mathbold{\sigma}}=\frac{1}{2}(\mathbold{\sigma}^{(1)}
-\mathbold{\sigma}^{(2)})$, instead of the individual
spin operators $\mathbold{\sigma}^{(i)},~i=1,2$. At the next step, one uses the known
values of matrix elements of the operators ${\bf Sn}$, $\mathbold{\sigma} {\bf n}$, and
the tensor operators
\eq\label{tensor}
S_{12}=3(\mathbold{\sigma}^{(1)}{\bf n})(\mathbold{\sigma}^{(2)}{\bf n})-(\mathbold{\sigma}^{(1)}\mathbold{\sigma}^{(2)})
=6({\bf Sn})^2-2{\bf S}^2\, ,\quad\quad
(\mathbold{\sigma}^{(1)}\mathbold{\sigma}^{(2)})=2{\bf S}^2-2
\en
between the different spin-angular momentum states
\eq\label{matrix_elements}
&&\langle LSJM_J|\pmatrix{{\bf S} {\bf n}\cr {\bf \mathbold{\sigma}} {\bf n}}
|L'S'J'M_{J'}\rangle= \delta_{JJ'}
\delta_{M_J M_{J'}} \langle L||{\bf n}||L'\rangle  
\langle S||\pmatrix{{\bf S}\cr {\bf \mathbold{\sigma}}}||S'\rangle\times
\nonumber\\[2mm]
&&\hspace*{4.cm}\times\, W(LL'SS';1J) (-1)^{S'+L-J}\, ,
\nonumber\\[2mm]
&&\langle L||{\bf n}||L'\rangle =\sqrt{2L'+1}\,\,\langle L'100|L0\rangle,
\nonumber\\[2mm]
&&\langle S||{\bf S}||S'\rangle=\delta_{SS'} \sqrt{S(S+1)(2S+1)}\, ,
\nonumber\\[2mm]
&&\langle S||{\bf \mathbold{\sigma}} ||S'\rangle=(-1)^{S+1} \sqrt{3}
 \delta_{S|S'-1|}
\nonumber\\[2mm]
&&\langle LSJM_{J}|\hat S_{12}|L'S'J'M_{J'}\rangle=
\delta_{JJ'} \delta_{M_J M_J'} \delta_{SS'}\,\biggl\{
\frac{\langle LSJ|| \hat S_{12} ||L'S'J'\rangle} {\sqrt{2J+1}}=
(-1)^{L-L'}\delta_{S1} \times
\nonumber\\[2mm]
&&\hspace*{4.cm}\times\, \sqrt{120(2L'+1)} <L'200|L0>W(LJ21;1L')\biggr\}\, ,
\en
where $W$ stand for the conventional Racah coefficients.  
The bound-state equations for the radial wave functions 
$R^{(\alpha_1\alpha_2)}_{LSJ}(p)$ are then obtained straightforwardly
\eq\label{PW-wf-1}
&&\bigl[\, M_B-(\alpha_1w_1+\alpha_2w_2)\,\bigr]\,
R^{(\alpha_1\alpha_2)}_{J{\tiny\pmatrix{0\cr 1}}J}(p)=
A^{(\alpha_1\alpha_2)}(M_B;p)\sum_{\alpha_1\alpha_2}\int_0^\infty
p'^2dp'\times
\nonumber\\[2mm]
&\times&\biggl\{\biggl[\biggl( (
{\cal N}_{12}^{(\alpha_1\alpha_2)}(p)
{\cal N}_{12}^{(\alpha_1'\alpha_2')}(p')
+\alpha_1\alpha_2\alpha_1'\alpha_2'
{\cal N}_{12}^{(-\alpha_1-\alpha_2)}(p)
{\cal N}_{12}^{(-\alpha_1'-\alpha_2')}(p'))
V_1^{J}(p,p')
\nonumber\\[2mm]
&+&(\alpha_1\alpha_1'{\cal N}_{12}^{(-\alpha_1\alpha_2)}(p)
  {\cal N}_{12}^{(-\alpha_1'\alpha_2')}(p')
+\alpha_2\alpha_2'{\cal N}_{12}^{(\alpha_1-\alpha_2)}(p)
  {\cal N}_{12}^{(\alpha_1'-\alpha_2')}(p'))
V_{2\oplus J}^{\tiny\pmatrix{0\cr 1}}(x;p,p')\biggr)\, 
R^{(\alpha_1'\alpha_2')}_{J{\tiny\pmatrix{0\cr 1}}J}(p')
\nonumber\\[2mm]
&-&(\alpha_1\alpha_1'{\cal N}_{12}^{(-\alpha_1\alpha_2)}(p)
  {\cal N}_{12}^{(-\alpha_1'\alpha_2')}(p')
-\alpha_2\alpha_2'{\cal N}_{12}^{(\alpha_1-\alpha_2)}(p)
  {\cal N}_{12}^{(\alpha_1'-\alpha_2')}(p'))
V_{2\ominus J}(x;p,p')\biggr)\, 
R^{(\alpha_1'\alpha_2')}_{J{\tiny\pmatrix{1\cr 0}}J}(p')\biggr]
\nonumber\\[2mm]
&+&\biggl[\biggl( (
{\cal N}_{12}^{(\alpha_1\alpha_2)}(p)
{\cal N}_{12}^{(\alpha_1'\alpha_2')}(p')
+\alpha_1\alpha_2\alpha_1'\alpha_2'
{\cal N}_{12}^{(-\alpha_1-\alpha_2)}(p)
{\cal N}_{12}^{(-\alpha_1'-\alpha_2')}(p'))
\nonumber\\[2mm]
&\pm& (\alpha_1\alpha_2{\cal N}_{12}^{(-\alpha_1-\alpha_2)}(p)
+\alpha_1'\alpha_2'{\cal N}_{12}^{(-\alpha_1'-\alpha_2')}(p')))\,
V_{\rm T}^J(p,p')
\nonumber\\[2mm]
&-&(\alpha_1\alpha_1'{\cal N}_{12}^{(-\alpha_1\alpha_2)}(p)
{\cal N}_{12}^{(-\alpha_1'\alpha_2')}(p')
+\alpha_2\alpha_2'{\cal N}_{12}^{(\alpha_1-\alpha_2)}(p)
 {\cal N}_{12}^{(\alpha_1'-\alpha_2')}(p')
\nonumber\\[2mm]
&\pm& (\alpha_1\alpha_2{\cal N}_{12}^{(-\alpha_1-\alpha_2)}(p)
+\alpha_1'\alpha_2'{\cal N}_{12}^{(-\alpha_1'-\alpha_2')}(p')))\,
V_{{\rm T}\oplus J}^{\tiny\pmatrix{0\cr 1}}(p,p')\biggr)\,
R^{(\alpha_1'\alpha_2')}_{J{\tiny\pmatrix{0\cr 1}}J}(p')
\nonumber\\[2mm]
&+&((\alpha_1\alpha_1'{\cal N}_{12}^{(-\alpha_1\alpha_2)}(p)
{\cal N}_{12}^{(-\alpha_1'\alpha_2')}(p')
-\alpha_2\alpha_2'{\cal N}_{12}^{(\alpha_1-\alpha_2)}(p)
 {\cal N}_{12}^{(\alpha_1'-\alpha_2')}(p')
\nonumber\\[2mm]
&\mp& (\alpha_1\alpha_2{\cal N}_{12}^{(-\alpha_1-\alpha_2)}(p)
-\alpha_1'\alpha_2'{\cal N}_{12}^{(-\alpha_1'-\alpha_2')}(p')))\,
V_{{\rm T}\ominus J}(p,p')\biggl)\,
R^{(\alpha_1'\alpha_2')}_{J{\tiny\pmatrix{1\cr 0}}J}(p')
\biggl]\biggl\}\, ,
\en

\eq\label{PW-wf-2}
&&\bigl[\, M_B-(\alpha_1w_1+\alpha_2w_2)\,\bigr]\,
R^{(\alpha_1\alpha_2)}_{J\pm 1 1J}(p)=
A^{(\alpha_1\alpha_2)}(M_B;p)\sum_{\alpha_1\alpha_2}\int_0^\infty
p'^2dp'\times
\nonumber\\[2mm]
&\times&\biggl\{\biggl[\biggl(
{\cal N}_{12}^{(\alpha_1\alpha_2)}(p)
{\cal N}_{12}^{(\alpha_1'\alpha_2')}(p')
V_1^{J\mp 1}(p,p')
+\alpha_1\alpha_2\alpha_1'\alpha_2'
{\cal N}_{12}^{(-\alpha_1-\alpha_2)}(p)
{\cal N}_{12}^{(-\alpha_1'-\alpha_2')}(p')
V_{1(J\pm 1)J}(p,p')
\nonumber\\[2mm]
&+&(\alpha_1\alpha_1'{\cal N}_{12}^{(-\alpha_1\alpha_2)}(p)
  {\cal N}_{12}^{(-\alpha_1'\alpha_2')}(p')
+\alpha_2\alpha_2'{\cal N}_{12}^{(\alpha_1-\alpha_2)}(p)
  {\cal N}_{12}^{(\alpha_1'-\alpha_2')}(p'))
V_2^J(x;p,p')\biggr)\, 
R^{(\alpha_1'\alpha_2')}_{J\pm 1 1J}(p')
\nonumber\\[2mm]
&+&\biggl(\alpha_1\alpha_2\alpha_1'\alpha_2'
{\cal N}_{12}^{(-\alpha_1-\alpha_2)}(p)
{\cal N}_{12}^{(-\alpha_1'-\alpha_2')}(p')
\frac{2}{2J+1}\, V_{1\ominus J}(p,p')\biggr)\,
R^{(\alpha_1'\alpha_2')}_{J\mp 1 1J}(p')\biggr]
\nonumber\\[2mm]
&+&\biggl[\biggl( (
{\cal N}_{12}^{(\alpha_1\alpha_2)}(p)
{\cal N}_{12}^{(\alpha_1'\alpha_2')}(p')
\pm (\alpha_1\alpha_2{\cal N}_{12}^{(-\alpha_1-\alpha_2)}(p)
+\alpha_1'\alpha_2'{\cal N}_{12}^{(-\alpha_1'-\alpha_2')}(p'))
\frac{1}{2J+1})\, V_{\rm T}^{J\pm 1}(p,p')
\nonumber\\[2mm]
&+&\alpha_1\alpha_2\alpha_1'\alpha_2'
{\cal N}_{12}^{(-\alpha_1-\alpha_2)}(p)
{\cal N}_{12}^{(-\alpha_1'-\alpha_2')}(p')
V_{{\rm T}(J\pm 1)J}(p,p')
-(\alpha_1\alpha_1'{\cal N}_{12}^{(-\alpha_1\alpha_2)}(p)
  {\cal N}_{12}^{(-\alpha_1'\alpha_2')}(p')
\nonumber\\[2mm]
&+&\alpha_2\alpha_2'{\cal N}_{12}^{(\alpha_1-\alpha_2)}(p)
  {\cal N}_{12}^{(\alpha_1'-\alpha_2')}(p')
\pm (\alpha_1\alpha_2'{\cal N}_{12}^{(-\alpha_1\alpha_2)}(p)
{\cal N}_{12}^{(\alpha_1'-\alpha_2')}(p')
\nonumber\\[2mm]
&+&\alpha_2\alpha_1'{\cal N}_{12}^{(\alpha_1-\alpha_2)}(p)
{\cal N}_{12}^{(-\alpha_1'\alpha_2')}(p'))\frac{1}{2J+1})
V_{\rm T}^J(p,p')\biggr)
R^{(\alpha_1'\alpha_2')}_{J\pm 1 1 J}(p')
\nonumber\\[2mm]
&+&\biggl(
\alpha_1\alpha_2\alpha_1'\alpha_2'
{\cal N}_{12}^{(-\alpha_1-\alpha_2)}(p)
{\cal N}_{12}^{(-\alpha_1'-\alpha_2')}(p')
\frac{2}{2J+1}V_{{\rm T}\ominus J}(p,p')
-(\alpha_1\alpha_1'{\cal N}_{12}^{(-\alpha_1\alpha_2)}(p)
  {\cal N}_{12}^{(-\alpha_1'\alpha_2')}(p')
\nonumber\\[2mm]
&+&\alpha_2\alpha_2'{\cal N}_{12}^{(\alpha_1-\alpha_2)}(p)
  {\cal N}_{12}^{(\alpha_1'-\alpha_2')}(p'))
\frac{2\sqrt{J(J+1)}}{2J+1}\,
V_{\rm T}^{J\mp 1}(p,p')
\mp(\alpha_1\alpha_2'{\cal N}_{12}^{(-\alpha_1\alpha_2)}(p)
{\cal N}_{12}^{(\alpha_1'-\alpha_2')}(p')
\nonumber\\[2mm]
&+&\alpha_2\alpha_1'{\cal N}_{12}^{(\alpha_1-\alpha_2)}(p)
{\cal N}_{12}^{(-\alpha_1'\alpha_2')}(p'))
\frac{2\sqrt{J(J+1)}}{2J+1}\,
V_{\rm T}^J(p,p')\biggr)\,
R^{(\alpha_1'\alpha_2')}_{J\mp 1 1 J}(p')
\biggr]\biggr\}\, ,
\en
where
\eq\label{Vplus}
V_{{\rm A}\oplus J}^{\tiny\pmatrix{0\cr 1}}(p,p')
&=&\frac{1}{2J+1}\biggl[
\pmatrix{J\cr J+1}V_{\rm A}^{J-1}(p,p')
+\pmatrix{J+1\cr J}V_{\rm A}^{J+1}(p,p')\biggr]\, ,
\nonumber\\[2mm]
V_{{\rm A}\ominus J}(p,p')&=&\frac{\sqrt{J(J+1)}}{2J+1}\biggl[
V_{\rm A}^{J-1}(p,p')-V_{\rm A}^{J+1}(p,p')\biggr]\, ,
\nonumber\\[2mm]
V_{{\rm A}(J\pm 1)J}(p,p')&=&
\frac{1}{(2J+1)^2}\biggl[
V_{\rm A}^{J\pm 1}(p,p')+4J(J+1)V_{\rm A}^{J\pm 1})(p,p')
\biggr]\, ,
\quad\quad {\rm A}=1,2,{\rm T}\, .
\en

The normalization condition in terms of radial wave functions
has a particularly simple form
\eq\label{PW-norm}
\sum_{\alpha_1\alpha_2}\int_0^\infty\frac{p^2dp}{(2\pi)^3}\,
f_{12}^{(\alpha_1\alpha_2)}(M_B;p)\,\,
\biggl[ R_{LSJ}^{(\alpha_1\alpha_2)}(p)\biggr]^2=2M_B\, .
\en
  
\subsection{Dynamical input}

For solving the bound-state equation, one needs further to specify
the inter-quark potentials $V_{\rm OG},~V_{\rm C},~V_{\rm T}$, introduced
above. Let us start from the confining part of the potential.
It is believed that the explicit form of this potential (i.e. its
dependence on the inter-quark distance) is in principle, derivable from
QCD. At present, however, the only tangible theoretical constraint on
the form of this potential is the linear growth at large distances
obtained within the quenched lattice QCD~\cite{Wilson}. Less compelling
arguments based on the background field technique, were provided to
justify the harmonic oscillator-type ($\sim r^2$) behavior of the
confining potential at small distances. With no rigorous solution of
the problem in sight, one may use the potential that interpolates
between the ``known'' behavior of the potential in different limiting
situations~\cite{Mittal,Gupta} (for a slightly modified version,
see~\cite{AChKR}) 
\eq\label{interpolation}
V_{\rm C}(r)=\frac{4}{3}\, \alpha_{\rm S}(m_{12}^2)\,
\biggl(\frac{\mu_{12}\omega_0^2 r^2}{2\sqrt{1+A_0m_1m_2 r^2}}
-V_0\biggr)\, , 
\en
\eq\label{V-definitions}
\alpha_{\rm S}=\frac{12\pi}{33-2n_f}\biggl(\ln\frac{Q^2}{\Lambda_{QCD}^2}
\biggr)^{-1}\, ,\quad\quad
m_{12}=m_1+m_2\, ,\quad\quad
\mu_{12}=\frac{m_1m_2}{m_{12}}\, ,
\en
where $Q^2$ is the momentum transfer squared, and the factor $\frac{4}{3}$
comes from the color-dependent part of the $q\bar q$ interaction.
$n_f$ is the number of flavors ($n_f=3$ for $u,d,s$ quarks, $n_f=4$
for $u,d,s,c$ quarks, $n_f=5$ for $u,d,s,c,b$
quarks). $\omega_0,~V_0,~A_0,~\Lambda_{QCD}$ are considered to be the
free parameters of the model. The potential given by
Eq.~(\ref{interpolation}), effectively reduces to the harmonic
oscillator potential for the light quarks $u,d,s$, and to the linear
potential for the heavy $b,c$ quarks, that meets our expectations.
In these limiting cases, the potential takes the form
\eq\label{limiting}
{\rm LINEAR}\quad &:&\quad
V_{\rm C}(r)=\frac{4}{3}\alpha_{\rm S}(m_{12}^2)
\biggl(\frac{\omega_0^2}{2m_{12}}\sqrt{\frac{m_1m_2}{A_0}}\, r-V_0\biggr)
 \equiv a_1+b_1 r\, ,
\nonumber\\[2mm]
{\rm HARMONIC}\quad &:&\quad
V_{\rm C}(r)=\frac{4}{3}\alpha_{\rm S}(m_{12}^2)
\biggl(\frac{\mu_{12}\omega_0^2}{2}\, r^2-V_0\biggr)\equiv
a_2+b_2 r^2\, .
\en

The one-gluon exchange potential is given by the standard
expression~\cite{AChKR} 
\eq\label{OG}
V_{\rm OG}(r)=-\frac{4}{3}\frac{\alpha_{\rm S}(m_{12}^2)}{r}
\equiv b_{-1} r^{-1}\, .
\en

Noting that
\eq\label{derivative}
r^n=\lim_{\eta\rightarrow 0}(-)^n
\frac{\partial^n}{\partial\eta^n}\biggl(
\frac{{\rm e}^{-\eta r}}{r}\biggr)\, ,
\en
one can rewrite the potentials in the momentum space
\eq
\label{lin-momentum}
{\rm LINEAR}\quad &:&\quad
V_{\rm C}({\bf p}-{\bf p}')=
a_1(2\pi)^3\delta^3({\bf p}-{\bf p}')+
b_1\lim_{\eta\rightarrow 0}\frac{\partial^2}{\partial\eta^2}
\biggl(\frac{4\pi}{|{\bf p}-{\bf p}'|^2}\biggr)\, ,
\\[2mm]
\label{har-momentum}
{\rm HARMONIC}\quad &:&\quad
V_{\rm C}({\bf p}-{\bf p}')=
\biggl( a_2-b_2\triangle_{{\bf p}'}\biggr)
(2\pi)^3\delta^3({\bf p}-{\bf p}')\, ,
\\[2mm]
\label{og-momentum}
{\rm ONE-GLUON}\quad &:&\quad
V_{\rm C}({\bf p}-{\bf p}')=
b_{-1}\,\frac{4\pi}{|{\bf p}-{\bf p}'|^2}\, .
\en

In order to investigate the properties
of the $q\bar q$ bound systems, the linear potential
was used both in
the configuration space~\cite{Long,Lagae,Olsson,Tjon}, and in the 
momentum space~\cite{AChK-FBS,Vary,Parramore-1,Vary-1}.
In the latter case, a special numerical algorithm based on the
the regularization~(\ref{lin-momentum}), was
utilized~\cite{Vary,Vary-1}.
In Refs.~\cite{Resag-2,Resag-3,Resag-4,Zoller,Klempt},  
the matrix elements of $V_{\rm C}(r)$ were calculated in the
configuration-space basis, in order to encompass the difficulties
related to the singular character of the linear potential in the
momentum space.

The investigation of the $q\bar q$ systems in the framework of Salpeter
equation was carried out~\cite{ChK-Czech,ChK,AChK-Yad,Lagae,AChKR},
using the harmonic confining potential.
MW, CJ and MNK equations with the harmonic confinement were considered
in Refs.~\cite{PLB,BKR}.

Some mathematical problems arise if the one-gluon exchange potential
with the {\it fixed} coupling constant $b_{-1}$ is used for the
calculation of the characteristics of $q\bar q$ bound systems.
Namely, as it was shown in Ref.~\cite{Resag-3,Zoller}, 
in this case the Salpeter wave
function is divergent at $r\rightarrow 0$. For the running coupling
constant this divergence is less pronounced but still present - now,
the problem occurs in the decay observables which depend on the value
of the wave function at $r\rightarrow 0$. In order to cure this
divergence, in Refs.~\cite{Resag-3,Zoller} the following
regularization was proposed
\eq\label{OG-regularization}
V_{\rm OG}(r)=-\frac{4}{3}\frac{\alpha_{\rm S}(r)}{r}\, ,
\quad\quad && {\rm for~}r>r_0\, ,
\nonumber\\[2mm]  
V_{\rm OG}(r)=a_g r^2+b_g\, ,
\quad\quad && {\rm for~}r<r_0\, ,
\en
where
\eq\label{alphaS}
\alpha_{\rm S}(r)=\frac{A}
{2\ln({\rm e}^{-(\gamma+\mu a)}/a+
{\rm e}^{A/(2\alpha_{sat})})}\biggl[
1-B\,\,\frac{\ln(2\ln({\rm e}^{-\tilde\mu a}/a+{\rm e}^{1/2}))}
{2\ln({\rm e}^{-\tilde\mu a}/a+{\rm e}^{B/2})}\biggr]\, ,
\en
where $a=\Lambda_{QCD}r$, $\gamma=0.577215\cdots$ is the
Euler-Mascheroni constant, and $\alpha_{sat}=0.4$, $\mu=4$,
$\tilde\mu=20$. Further,
\eq\label{alphaS-parameters}
A=\frac{12\pi}{33-2n_f}\, ,\quad\quad
B=\frac{6(53-19n_f)}{(33-2n_f)^2}\, .
\en
Note, that in Ref.~\cite{Tjon}, the choice $B=0$ is adopted.
The constants $a_g$ and $b_g$ from Eq.~(\ref{OG-regularization})
are determined from matching of the potential and its first derivative
at $r=r_0$. It turns out that the dependence of the $q\bar q$ system
mass spectrum on the regularization parameter $r_0$ is very weak
provided the latter is chosen to be sufficiently small.

\subsection{t'Hooft interaction}
 
The t'Hooft interaction is used in the form suggested in
Ref.~\cite{Resag-2}. The point-like potential in the configuration
space would lead to the divergences. For this reason, the following
regularization of the potential is considered
\eq\label{tHooft-reg}
V_{\rm T}(r)\rightarrow 4\hat g V_{{\rm T},reg}(r;\Lambda)\, ,\quad
V_{{\rm T},reg}(r;\Lambda)=\frac{1}{(\Lambda\sqrt{\pi})^3}\,
\exp\biggl(-\frac{r^2}{\Lambda^2}\biggr)\, .
\en
In the momentum space, we have
\eq
V_{{\rm T},reg}({\bf p}-{\bf p}';\Lambda)=
\exp\biggl(-\frac{\Lambda^2({\bf p}-{\bf p}')^2}{4}\biggr)\, .
\en 
Now, using the following representation of the Legendre polynomials
\eq\label{Legendre}
P_n(z)=\frac{1}{2^n n!}\frac{d^n}{dz^n}\,(z^2-1)^n\, ,
\en
with the use of the identity
\eq\label{Legendre-identity}
\int {\rm e}^{az} z^n dz=\frac{\partial^n}{\partial a^n}
\int {\rm e}^{az} dz\, ,
\en
after the partial-wave expansion of the t'Hooft potential we obtain
\eq\label{tHooft-partial}
\lim_{\Lambda\rightarrow 0}
V_{{\rm T},reg}^{\bar L}(p,p';\Lambda)=
\delta_{\bar L0}
V_{{\rm T},reg}^{0}(p,p';0)\, ,
\en
that reflects the point-like character of the t'Hooft interaction.
Here,
\eq\label{tHooft-0}
V_{{\rm T},reg}^{0}(p,p';\Lambda)=
\frac{\exp\biggl(-\frac{\Lambda^2(p^2+{p'}^2)}{4}\biggr)}{4\pi^2}\,\,
\frac{4}{\Lambda^2pp'}\,\,
\sinh\frac{\Lambda^2pp'}{2}\, .
\en
In accordance with the Eq.~(\ref{tHooft-partial}),
all partial waves except $\bar L=0$ in the partial-wave
expansion of the t'Hooft potential are neglected 
even at nonzero $\Lambda$.

As it was mentioned above, the t'Hooft interaction was introduced in
order to provide the mass splitting between the pseudoscalar and
vector octets within the framework of the constituent quark model,
which in QCD is due to the spontaneous breaking of chiral
symmetry. The quantity $\hat g$ that appears in
Eq.~(\ref{tHooft-reg}), is the matrix in the flavor space. The matrix
elements of this matrix between various meson states
\eq\label{mesons}
\pi,~\rho,~K,~\omega=\eta_n=\frac{1}{\sqrt{2}}(u\bar u+d\bar d)\, ,~
\Phi=\eta_s=s\bar s\, ,
\en
are~\cite{Resag-2} 
\eq\label{matrix-elements}
&&\langle\pi|\hat g|\pi\rangle=-g=\langle\rho|\hat g|\rho\rangle\, ,\quad
\langle K|\hat g|K\rangle=-g'\, ,
\nonumber\\[2mm]
&&\langle\eta_s|\hat g|\eta_s\rangle=0\, ,\quad
\langle\eta_n|\hat g|\eta_n\rangle=g\, ,\quad
\langle\eta_n|\hat g|\eta_s\rangle=\sqrt{2}g'
\langle\eta_s|\hat g|\eta_n\rangle\, ,
\en
where $g$ and $g'$ are two independent coupling constants that are
considered to be the free parameters of the model. 

The $\eta$ and $\eta'$ mesons are the superpositions of $\eta_n$ and
$\eta_s$. In order to take the mixing into account, we introduce the
matrix notations
\eq\label{matrix-notations}
&&\tilde\Phi_{M_B}({\bf p})=\pmatrix{\tilde\Phi_{n,M_B}({\bf p})\cr
\tilde\Phi_{s,M_B}({\bf p})}\, ,\quad
\tilde G_0(M_B;{\bf p})=\pmatrix{\tilde G_{n,0}(M_B;{\bf p})&0\cr
0&\tilde G_{s,0}(M_B;{\bf p})}\, ,
\nonumber\\[2mm]
&&V_{\rm A}({\bf p},{\bf p}')=
\pmatrix{V_{n,{\rm A}}({\bf p},{\bf p}')&0\cr
0&V_{n,{\rm A}}({\bf p},{\bf p}')}\, ,~~{\rm A}={\rm C,OG}\, ,\quad
\nonumber\\[2mm]
&&V_{\rm T}({\bf p},{\bf p}')=
\pmatrix{V_{nn,{\rm T}}({\bf p},{\bf p}')&
V_{ns,{\rm T}}({\bf p},{\bf p}')\cr
V_{sn,{\rm T}}({\bf p},{\bf p}')&
V_{ss,{\rm T}}({\bf p},{\bf p}')}\, .
\en
The radial wave functions $R_{f,000}^{(\alpha_1\alpha_2)}(p),~f=n,s$
describing the $\eta$ and $\eta'$ mesons, obey the following system of
equations
\eq\label{eq-tHooft}
&&\bigl[\, M_{B}-(\alpha_1w_1+\alpha_2w_2)\bigr]_f
R_{f,000}^{(\alpha_1\alpha_2)}(p)=
A_f^{(\alpha_1\alpha_2)}(M_B;p)\sum_{\alpha_1'\alpha_2'}
\int_0^\infty {p'}^2dp'\times
\nonumber\\[2mm]
&\times&\biggl\{\biggl[ (
{\cal N}_{f,12}^{(\alpha_1\alpha_2)}(p)
{\cal N}_{f,12}^{(\alpha_1'\alpha_2')}(p')
+\alpha_1\alpha_2\alpha_1'\alpha_2'
{\cal N}_{f,12}^{(-\alpha_1-\alpha_2)}(p)
{\cal N}_{f,12}^{(-\alpha_1'-\alpha_2')}(p'))
V_1^0(p,p')
\nonumber\\[2mm]
&+&(\alpha_1\alpha_1'{\cal N}_{f,12}^{(-\alpha_1\alpha_2)}(p)
{\cal N}_{f,12}^{(-\alpha_1'\alpha_2')}(p')
+\alpha_2\alpha_2'{\cal N}_{f,12}^{(\alpha_1-\alpha_2)}(p)
{\cal N}_{f,12}^{(\alpha_1'-\alpha_2')}(p'))
V_2^1(p,p')\biggr]\,
R_{f,000}^{(\alpha_1'\alpha_2')}(p')
\nonumber\\[2mm]
&+&\sum_{f'}\biggl[ (
{\cal N}_{f,12}^{(\alpha_1\alpha_2)}(p)
{\cal N}_{f',12}^{(\alpha_1'\alpha_2')}(p')
+\alpha_1\alpha_2\alpha_1'\alpha_2'
{\cal N}_{f,12}^{(-\alpha_1-\alpha_2)}(p)
{\cal N}_{f',12}^{(-\alpha_1'-\alpha_2')}(p')
\nonumber\\[2mm]
&+&\alpha_1\alpha_2{\cal N}_{f,12}^{(-\alpha_1-\alpha_2)}(p)
+\alpha_1'\alpha_2'{\cal N}_{f',12}^{(-\alpha_1'-\alpha_2')}(p'))
V_{{\rm T},reg}^{0}(p,p';\Lambda)
\langle\eta_f|4\hat g|\eta_{f'}\rangle\biggr]
R_{f',000}^{(\alpha_1'\alpha_2')}(p')\biggr\}\, .
\en
The functions $R_{f,000}^{(\alpha_1\alpha_2)}(p),~f=n,s$ satisfy the
normalization condition
\eq\label{norm-tHooft}
\int_0^\infty\frac{p^2dp}{(2\pi)^3}\,\sum_{\alpha_1\alpha_2}
\biggl[ f_{n,12}^{(\alpha_1\alpha_2)}(M_B,p) 
\biggl( R_{n,000}^{(\alpha_1\alpha_2)}(p)\biggr)^2
+f_{s,12}^{(\alpha_1\alpha_2)}(M_B,p) 
\biggl( R_{s,000}^{(\alpha_1\alpha_2)}(p)\biggr)^2\biggr]=2M_B\, ,
\en
where $M_B$ is either $M_\eta$ or $M_{\eta'}$.

The equations for other mesonic states can be obtained, replacing
$\langle\eta_f|4\hat g|\eta_{f'}\rangle,~f,f'=n,s$ by the
corresponding matrix elements from Eq.~(\ref{matrix-elements}).

Note that the mixing in $\Phi-\omega$ and $\eta-\eta'$ systems has
been recently also investigated in Refs.~\cite{Shakin-1} within the
Nambu-Jona-Lasinio (NJL) model, with an account of the relativistic confinement
potential (Lorentz vector structure only) and the t'Hooft interaction.

\subsection{Solution of the equations} 

One has to specify
the numerical procedure for the solution of the system of radial
equations~(\ref{PW-wf-1})-(\ref{PW-wf-2}). 
A possible algorithm looks as follows.
One chooses the known basis functions denoted by $R_{nL}(p)$.
The radial wave functions are expanded in the linear combinations of
the basis functions
\eq\label{basis}
R_{LSJ}^{(\alpha_1\alpha_2)}(p)=\sqrt{2M_B(2\pi)^3}\, 
\bar R_{LSJ}^{(\alpha_1\alpha_2)}(p)
=\sqrt{2M_B(2\pi)^3}\, 
\sum_{n=0}^\infty c_{nLSJ}^{(\alpha_1\alpha_2)}R_{nL}(p)\, ,
\en
where $c_{nLSJ}^{(\alpha_1\alpha_2)}$ are the coefficients of the expansion.
The integral equation for the radial wave functions is then
transformed into the system of linear equations for these
coefficients. If the truncation is carried out, the finite system of
equations is obtained that can be solved by using conventional
numerical methods. The convergence of the whole
procedure, with more terms taken into account in the
expansion~(\ref{basis}), depends on the successful choice of the
basis. In
Refs.~\cite{Resag-1,Resag-2,Resag-3,Resag-4,Zoller,Olsson,Klempt,Munz},
where the linear confining potential is assumed, the basis functions
are chosen in the following manner
\eq\label{Laguerre}
R_{nL}(y)=N_{nL}y^LL_n^{2L+2}(y){\rm e}^{-y/2}\, ,\quad\quad 
y=\beta p\, ,
\en
where $L_n^{2L+2}(y)$ are the Laguerre polynomials, and $\beta$ is the
free parameter. In Refs.~\cite{Long,Parramore,Parramore-1}, the
non-relativistic oscillator wave functions (again containing the free
parameter), were used in spite of the fact that the linear confining
potential was assumed. In
Refs.~\cite{PLB,ChK-Czech,ChK,AChK-Yad,AChK-FBS,BKR}, 
the same basis functions were used, but without the free parameter,
due to the fact that the confining potential was taken in the harmonic
form, with the parameters already fixed. Finally, in
Ref.~\cite{AChKR}, the harmonic oscillator basis was used, whereas the
confining potential had the general form given by Eq.~(\ref{interpolation}).

To clarify the choice of the basis functions, let us consider the
non-relativistic limit of the equations~(\ref{frequency-wf-3}). 
In this limit,
one can replace $\gamma_0\rightarrow 1$, $\gamma_5\rightarrow 0$,
$\mathbold{\gamma}\rightarrow 0$. Consequently,
\eq\label{V-nonrel}
V\rightarrow V_{\rm OG}+V_{\rm C}+V_{\rm T}\, .
\en
Further, to derive the non-relativistic limit of the equations, we
expand the kinetic term $\alpha_1w_1+\alpha_2w_2$ in
Eq.~(\ref{frequency-wf-3}), retaining terms up to and including
$O({\bf p}^2/m_i^2)$. In the right-hand side of this equation, the
function $A^{(\alpha_1\alpha_2)}(M_B;p)$ can be replaced by its value
at $p=0$. In a result, we obtain
\eq\label{NR-limit-mix}
\tilde\Phi_{M_B;NR}^{(\pm\mp)}({\bf p})=0\, ,\quad\quad
\tilde\Phi_{M_B;NR}^{(--)}({\bf p})=0\, ,
\en
\eq\label{NR-limit-++}
\biggl[\,\varepsilon_B-\frac{{\bf p}^2}{2\mu_{12}}\,\biggr]\,
\tilde\Phi_{\varepsilon_B;NR}^{(++)}({\bf p})&=&
A^{(++)}(M_B;0)\,\int\frac{d^3{\bf p}'}{(2\pi)^3}\,
\bigl[\,V_{\rm OG}({\bf p}-{\bf p}')
\nonumber\\[2mm]
&+&V_{\rm C}({\bf p}-{\bf p}')
+V_{\rm T}({\bf p}-{\bf p}')\, \bigr]\,
\tilde\Phi_{\varepsilon_B;NR}^{(++)}({\bf p}')\, ,
\en
where $\varepsilon_B=M_B-m_{12}$, and
\eq\label{A++}
{\rm SAL,~GR,~MW}\quad &:&\quad
A^{(++)}(M_B;0)=1\, ,
\nonumber\\[2mm]
{\rm CJ}\quad &:&\quad
A^{(++)}(M_B;0)=\frac{1}{2}\, \biggl( 1+\frac{M_B}{m_{12}}\biggr)\, .
\en
The non-relativistic limit in the MNK version is more tricky.
For a general $M_B$, there emerges an arbitrary function of the ratio
$M_B/m_{12}$. However, if one uses the non-relativistic approximation
also for the bound-state mass $M_B=m_{12}$, then $A^{(++)}(M_B;0)=1$
for both the CJ and MNK versions. Below, we shall use this
approximation.

Since in the non-relativistic limit (see Eq.~(\ref{Pauli})) 
\eq\label{Pauli-NR}
\tilde\Phi_{\varepsilon_B;NR}^{(++)}({\bf p})
=\chi_{\varepsilon_B;NR}^{(++)}({\bf p})\, ,
\en
the non-relativistic limit of Eq.~(\ref{frequency-wf-3}) with the
harmonic confinement potential~(\ref{limiting}) only, is given by
\eq\label{osc-NR}
\biggl[\,\varepsilon_B-\frac{{\bf p}^2}{2\mu_{12}}+
\frac{4}{3}\,\alpha_{\rm S}(m_{12}^2)\biggl(
\frac{\mu_{12}\omega_0^2}{2}\,\triangle_{\bf p}+V_0\biggr)\biggr]\,
\chi_{\varepsilon_B;NR}^{(++)}({\bf p})=0\, .
\en
Performing the partial-wave expansion of Eq.~(\ref{osc-NR}), we obtain
the equation for the radial wave functions
\eq\label{osc-NR-pw}
\biggl[\,\frac{d^2}{dz^2}+\frac{2}{z}\,\frac{d}{dz}-\frac{L(L+1)}{z^2}
-z^2+\frac{2}{\omega_0}\sqrt{\frac{3}{4\alpha_{\rm S}(m_{12}^2)}}\,
\biggl(\varepsilon_B^{(n)}+\frac{4}{3}\,\alpha_{\rm S}(m_{12}^2)V_0\biggr)
\biggr]\, R_{L}(z)=0\, ,
\en
where $z=p/\bar p$, and 
$\bar p=\sqrt{\mu_{12}\omega_0\sqrt{\frac{4}{3}\alpha_{\rm S}(m_{12}^2)}}$.
The solutions of this equation with the energy spectrum
\eq\label{osc-NR-energy}
\varepsilon_B^{(n)}=-\frac{4}{3}\,\alpha_{\rm S}(m_{12}^2)V_0
+\sqrt{\frac{4}{3}\,\alpha_{\rm S}(m_{12}^2)}\,
\omega_0(2n+L+\frac{3}{2})\, ,
\en
are the well-known harmonic oscillator wave functions
\eq\label{osc-NR-wf}
&&R_{nL}(p)=\bar p^{-3/2}R_{nL}(z)\, ,\quad\quad
R_{nL}(z)=c_{nL}z^L\exp\biggl(-\frac{z^2}{2}\biggr)
~_1F_1(-n,L+\frac{3}{2},z^2)\, ,
\nonumber\\[2mm]
&&c_{nL}=\sqrt{\frac{2\Gamma(n+L+\frac{3}{2})}{\Gamma(n+1)}}\,
\frac{1}{\Gamma(L+\frac{3}{2})}\, ,
\en
where $~_1F_1$ denotes the confluent hypergeometric function.

The functions $R_{nL}(p)$ can be used as a basis for the expansion in
the general case~(\ref{basis}). The system of equations for the
coefficients is given by
\eq\label{system-lin}
M_Bc_{nLSJ}^{(\alpha_1\alpha_2)}=\sum_{\alpha_1'\alpha_2'}
\sum_{L'S'}\sum_{n'}
H_{LSJn;L'S'J'n'}^{(\alpha_1\alpha_2;\alpha_1'\alpha_2')}(M_B)
c_{n'L'S'J'}^{(\alpha_1'\alpha_2')}\, ,
\en
where the matrix $H(M_B)$ is given by the convolution of the potential
and various kinematic factors that appear in
Eqs.~(\ref{PW-wf-1})-(\ref{PW-wf-2}), with the wave functions of the
basis. From Eq.~(\ref{system-lin}) it is immediately seen that, in
general, the eigenvalue equation for $M_B$ is not a linear one, and
should be solved, e.g., by iterations.

In order to actually solve the system of equations~(\ref{system-lin}),
one has to truncate it at some fixed $n=N_{max}$. then,
$c^{(\alpha_1\alpha_2)}_{LSJn}$ are determined from the system of
$4(N_{max}+1)$ ($2(N_{max}+1)$ in Salpeter and Gross versions) linear
equations. This procedure determines the eigenvalue $M_B$ as well,
either directly, when the matrix $H(M_B)$ does not depend on $M_B$, or
by using the iterative procedure. Having solved the eigenvalue problem
at a fixed value of $N_{max}$, one has then to check the stability
with respect to the change of $N_{max}$ - if the calculated
eigenvalues do not converge with the increase of $N_{max}$, the
original system of integral equations is declared to have no
solutions.

Note that the system of equations~(\ref{system-lin}) is homogeneous in   
$c^{(\alpha_1\alpha_2)}_{LSJn}$. This means that the solution of the
eigenvalue problem determines these coefficients up to an overall
factor that can be fixed from the normalization condition
\eq\label{normalization-c}
\sum_{\alpha_1\alpha_2}\sum_{nn'}
c^{(\alpha_1\alpha_2)}_{LSJn}c^{(\alpha_1\alpha_2)}_{LSJn'}
\int_0^\infty p^2dp\, f_{12}^{(\alpha_1\alpha_2)}(M_B;p)\,
R_{nL}(p)R_{n'L}(p)=1\, .
\en
In the CJ and MNK versions, the function $f_{12}^{(--)}(M_B;p)$ has
the second-order pole, so in the normalization condition one
encounters singular integrals of the following type
\eq\label{singular}
I(x_0)=\int_0^\infty \frac{f(x)dx}{(x-x_0)^2}\, ,
\en
where $f(x)$ is the regular function that obeys the conditions
$f(0)=f(\infty)=0$. The integral in~(\ref{singular}) can be
regularized according to
\eq\label{regular}
\int_0^\infty \frac{f(x)dx}{(x-x_0)^2}=\int_0^{2x_0}
\frac{(f'(x)-f'(x_0))dx}{x-x_0}+\int_{2x_0}^\infty
\frac{f'(x)dx}{x-x_0}\, .
\en

The first question which one may be willing to investigate, is the
manifestation of the Lorentz structure of the confining interaction in
the bound-state mass spectrum, especially in the case of light quarks.
This question was addressed, e.g. in Ref.~\cite{ChK}, where the
scalar, timelike vector, and their equal-weight mixture were studied
on the basis of Salpeter equation (this corresponds to the choice
$x=0;1;0.5$ in Eq.~(\ref{V-static}), respectively). 
It was demonstrated that the stable solutions to the Salpeter equation
in the light quark sector
do not exist for the scalar confining potential $x=0$, and do exist
for $x=0.5$ and $x=1$. Further, in Ref.~\cite{ChK}, 
the structure $\gamma_\mu^{(1)}\otimes\gamma^{\mu(2)}$ was considered
as well - it was demonstrated that in the case the stable solutions do
not exist. In Ref.~\cite{AChK-Yad}, more general conclusion was
obtained - it was demonstrated that the stable solutions 
in the light quark sector exist for
any $x$ from the interval $0.5\leq x\leq 1$. This result was confirmed
in Refs.~\cite{Parramore,Olsson}. Further, in Ref.~\cite{AChK-Yad}, it
was shown that in the heavy quark sector nothing really depends on the
mixing parameter $x$ - the solutions exist everywhere and practically 
do not change when x varies in the whole interval $0\leq x\leq 1$.
This result is easy to understand. Indeed, the projection operator
$(\Lambda_{12}^{(++)}-\Lambda_{12}^{(--)})\gamma_0^{(1)}\otimes\gamma_0^{(2)}$,
that is present in the Salpeter equation, in the heavy quark limit is
equal to
$\frac{1}{2}(\gamma_0^{(1)}+\gamma_0^{(2)})\gamma_0^{(1)}\otimes\gamma_0^{(2)}$,
so that the confining interaction in this limit equals to
\eq\label{conf-hq}
&&(\Lambda_{12}^{(++)}-\Lambda_{12}^{(--)})\gamma_0^{(1)}\otimes\gamma_0^{(2)}\bigl[
x\gamma_0^{(1)}\otimes\gamma_0^{(2)}+(1-x)I^{(1)}\otimes I^{(2)}\bigr]
V_{\rm C}({\bf p}-{\bf p}')
\nonumber\\[2mm]
&&\rightarrow
\frac{1}{2}(\gamma_0^{(1)}+\gamma_0^{(2)})V_{\rm C}({\bf p}-{\bf p}')
\en
at $m_1,m_2\rightarrow\infty$, and does not depend on $x$ at all.
Note that in the literature we encounter the different choice of the
parameter $x$: $x=1$~\cite{Long,Lagae,Parramore},
$x=0.5$~\cite{Resag-3,Zoller,Klempt},
$x=0$~\cite{Lagae,Resag-2,Resag-3}.
Note also, that, as it was shown in Ref.~\cite{AChKR}, the
non-existence of the stable solutions at small $x$ in the light-quark
sector is related to the presence of the ``negative-energy'' component
in the Salpeter wave function.

The same question can be studied in other - GR, MW, CJ and MNK -
versions, that, unlike the Salpeter equation, have the correct
one-body limit. For the MW and CJ versions the investigations were
carried out in Ref.~\cite{Tjon}. Here, the problem was studied in
the configuration space, and for the confining potential the following
Lorentz structure was assumed: 
$V_{\rm C}({\bf p},{\bf p}')=\bigl[ \, 
x\gamma_\mu^{(1)}\otimes\gamma^{\mu(2)}+(1-x)I^{(1)}\otimes I^{(2)} 
\bigr]\, V_{\rm C}({\bf p}-{\bf p}')$ where
for $V_{\rm C}({\bf p}-{\bf p}')$ a linear form was chosen.
It was demonstrated that this potential should be ``more scalar than
vector'' in order to provide the existence of the stable
solutions. More detailed study of MW, CJ and MNK versions in the
momentum space was carried out in Refs.~\cite{PLB,BKR}, where the
harmonic confining potential was used, with the Lorentz structure
given by Eq.~(\ref{V-static}). The following states
$d\bar s:~^1S_0,^3S_1,^1P_1,^3P_0,^3P_1,^3P_2,^1P_2,^3D_1,^3D_3$,
$c\bar u~{\rm and}~c\bar s:~^1S_1,^1P_1,^3P_2$, were considered. 
It was demonstrated, that in all versions the solutions always exist
at $x=0$, whereas for $x=1$ for the majority of the states there is no
solution. This is just the opposite to the Salpeter equation case (see
above) - there, at $x=1$ there are the solutions, whereas at $x=0$, the
solutions for majority of states cease to exist. Put differently, the
existence/nonexistence of the solutions depends critically on the
value of $x$, and the criteria vary from version to version. In
addition, the criteria depend on the details of the potential - in
particular, the strength parameter $\omega_0$ introduced in
Eq.~(\ref{limiting}). Note that the instability mentioned, is now
caused by the admixture of the mixed $(+-),(-+)$ frequency components
in the bound-state wave function. One may look for the admissible
window in the parameter space, where the solutions of all versions
simultaneously exist and approximately coincide. In this way, one may
judge on the Lorentz structure - assuming that
the whole physical picture of the $q\bar q$ bound states based on the
3D reduction of the BS equation, is viable.     
From this study, one has to reject the MW version that poorly agrees
either with other versions or with data. Further, on the basis of
SAL, CJ and MNK versions, one can determine the acceptable interval
for the mixing parameter $x$: $0.3\leq x\leq 0.6$. 

Both the fine structure ($P$-wave splitting),
and the hyperfine structure ($^3S_1-^3D_1$ splitting) 
of the $q\bar q$ states depends on the value of the mixing parameter
$x$. As it was shown in Refs.~\cite{Resag-2,Parramore-1} on the basis
of Salpeter equation, the spin-orbit splitting in the light quarkonia
can only be described by the mixture of scalar and timelike vector
confinement. However, as was shown in Ref.~\cite{Parramore-1}, the
fine structure and the hyperfine structure can not be simultaneously
described by simply varying the value of the mixing parameter.
Finally, in Ref.~\cite{BKR}, more general - and pessimistic -
conclusion was drawn: neither of the versions - SAL, MW, CJ or MNK -
with the dynamical input specified above, does not describe even 
qualitative features of the whole mass spectrum of $q\bar q$ bound
states with $x$ inside the interval $0.3\leq x\leq 0.6$. 
Clearly, the problem calls for the further investigation.

\section{decays of the mesons in the CM frame}
\setcounter{equation}{0}

Further information about the bound $q\bar q$ systems may be gained, 
investigating their decays. Below, we consider exclusively the decays
that proceed into the CM frame of the bound state\footnote{
The treatment of the decays which can not be confined to the CM frame,
implies the specification of the Lorentz-transformation rules for the
instantaneous potentials and 3D wave functions. 
Due to the Lorentz covariance, the dependence on
the 0-th component of the relative momentum emerges into the
transformed wave functions, that renders the problem extremely
complicated, and the further assumptions are necessary. We do not
consider such processes here.}. These are:
the weak decays of the pseudoscalar mesons $P\rightarrow\mu\bar\nu$,
the leptonic decays of the neutral vector mesons $V\rightarrow^+e^-$,
and the two-photon decays $M\rightarrow\gamma\gamma$. The
corresponding characteristics are: the weak decay constant $f_P$, the
leptonic decay width $\Gamma(V\rightarrow e^+e^-)$ (or, the leptonic constant
$f_V$), and the two-photon decay width $\Gamma(M\rightarrow\gamma\gamma)$.  
   
The expressions for the quantities $f_P$ and $\Gamma(V\rightarrow
e^+e^-)$ were obtained in Refs. \cite{ChK,Resag-2,Resag-3,Resag-4} in
the framework of Salpeter equation, directly in terms of
$\tilde\Phi^{(\pm)}({\bf p})=\tilde\Phi_{aa}({\bf p})\pm
\tilde\Phi_{bb}({\bf p})$, or
$\tilde\Psi_{aa}({\bf p})=\tilde\Phi^{(++)}({\bf p})$,
$\tilde\Psi_{bb}({\bf p})=\tilde\Phi^{(--)}({\bf p})$
(see above). In Ref.~\cite{BKR}, these quantities were evaluated in
the framework of SAL, CJ and MNK versions written in the form
(\ref{eq-Pauli})-(\ref{Vplus}), that corresponds to the representation
of the wave function in the form (\ref{Pauli})-(\ref{cal-N}). The main
conclusion that comes from this investigation, is that the results do
not depend much on the choice of the different 3D reduction scheme.
The quantity $\Gamma(M\rightarrow\gamma\gamma)$ was evaluated in
Refs.~\cite{Resag-3,Resag-4,Klempt,Munz} for the systems
($\pi,\eta,\eta'$). Below, we shall follow the derivation presented in
Ref.~\cite{BKR}.

For the calculation of the quantities listed above, we need the wave
functions $\tilde\Phi^{(\alpha_1\alpha_2)}_{LSJM_J}({\bf p})$ which,
according to Eq.~(\ref{Pauli}), is expressed via
$\tilde\chi^{(\alpha_1\alpha_2)}_{LSJM_J}({\bf p})$. The partial-wave
expansion for the components of the wave function reads
\eq\label{pw-ab}
\bigl[\tilde\Phi^{(\alpha_1\alpha_2)}_{LSJM_J}({\bf p})\bigr]_{aa}&=&
\langle{\bf n}|LSJM_J\rangle\, {\cal N}_{12}^{(\alpha_1\alpha_2)}(p)\,
R^{(\alpha_1\alpha_2)}_{LSJ}(p)\, ,
\nonumber\\[2mm]
\bigl[\tilde\Phi^{(\alpha_1\alpha_2)}_{LSJM_J}({\bf p})\bigr]_{ab}&=&
-({\bf s}{\bf n}-{\bf\mathbold{\sigma}}{\bf n})\, 
\langle{\bf n}|LSJM_J\rangle\,
\frac{\alpha_2p{\cal N}^{(\alpha_1\alpha_2)}_{12}(p)}{w_2+\alpha_2m_2}\,
R^{(\alpha_1\alpha_2)}_{LSJ}(p)\, ,
\nonumber\\[2mm]
\bigl[\tilde\Phi^{(\alpha_1\alpha_2)}_{LSJM_J}({\bf p})\bigr]_{ba}&=&
({\bf s}{\bf n}+{\bf\mathbold{\sigma}}{\bf n})\, 
\langle{\bf n}|LSJM_J\rangle\,
\frac{\alpha_1p{\cal N}^{(\alpha_1\alpha_2)}_{12}(p)}{w_1+\alpha_1m_1}\,
R^{(\alpha_1\alpha_2)}_{LSJ}(p)\, ,
\nonumber\\[2mm]
\bigl[\tilde\Phi^{(\alpha_1\alpha_2)}_{LSJM_J}({\bf p})\bigr]_{bb}&=&
-\frac{{\bf S}_{12}+{\bf \mathbold{\sigma}}^{(1)}{\bf\mathbold{\sigma}}^{(2)}}{3}\,
\langle{\bf n}|LSJM_J\rangle\,
\frac{\alpha_1\alpha_2p^2{\cal N}^{(\alpha_1\alpha_2)}_{12}(p)}
{(w_1+\alpha_1m_1)(w_2+\alpha_2m_2)}\,
R^{(\alpha_1\alpha_2)}_{LSJ}(p)\, ,
\en
where ${\bf S}=\frac{1}{2}({\bf\mathbold{\sigma}}^{(1)}+{\bf\mathbold{\sigma}}^{(2)})$,
 ${\bf\mathbold{\sigma}}=\frac{1}{2}({\bf\mathbold{\sigma}}^{(1)}-{\bf\mathbold{\sigma}}^{(2)})$,
and the operators ${\bf S}_{12}$ and ${\bf\mathbold{\sigma}}^{(1)}{\bf\mathbold{\sigma}}^{(2)}$
are given by Eq.~(\ref{tensor}).

Using now the identity which is valid for any operator $\hat{\bf O}$
\eq\label{O}
\hat{\bf O}({\bf\mathbold{\sigma}}^{(1)},{\bf\mathbold{\sigma}}^{(2)},{\bf n})\,
\langle{\bf n}|LSJM_J\rangle=
\sum_{L'S'J'M_{J'}}\,\langle{\bf n}|L'S'J'M_{J'}\rangle\,
\langle L'S'J'M_{J'}|\hat{\bf O}|LSJM_J\rangle\, ,
\en
and the well-known expressions for the matrix elements of the operators
${\bf S}{\bf n}$, ${\bf\mathbold{\sigma}}{\bf n}$, ${\bf S}_{12}$, from Eq.~(\ref{pw-ab})
it is straightforward to obtain
\eq\label{pw-ab-lsj}
\bigl[\tilde\Phi^{(\alpha_1\alpha_2)}_{LSJM_J}({\bf p})\bigr]_{aa}&=&
\langle{\bf n}|LSJM_J\rangle\, {\cal N}_{12}^{(\alpha_1\alpha_2)}(p)\,
R^{(\alpha_1\alpha_2)}_{LSJ}(p)\, ,
\nonumber\\[2mm]
\bigl[\tilde\Phi^{(\alpha_1\alpha_2)}_
{J{\tiny\pmatrix{0\cr1}}JM_J}({\bf p})\bigr]_{ab}&=&
\biggl[\mp\sqrt{\frac{{\footnotesize\pmatrix{J+1\cr J}}}{2J+1}}\,
\langle{\bf n}|J+1 1 J M_J\rangle
+\sqrt{\frac{{\footnotesize\pmatrix{J\cr J+1}}}{2J+1}}\,
\langle{\bf n}|J-1 1 J M_J\rangle\,\biggr]\times
\nonumber\\[2mm]
&&\times\alpha_2{\cal N}_{12}^{(\alpha_1-\alpha_2)}(p)\,
R^{(\alpha_1\alpha_2)}_{J{\tiny\pmatrix{0\cr 1}}J}(p)\, ,
\nonumber\\[2mm]
\bigl[\tilde\Phi^{(\alpha_1\alpha_2)}_
{J\pm11JM_J}({\bf p})\bigr]_{ab}&=&
\biggl[\sqrt{\frac{{\footnotesize\pmatrix{J\cr J+1}}}{2J+1}}\,
\langle{\bf n}|J 1 J M_J\rangle
\mp\sqrt{\frac{{\footnotesize\pmatrix{J+1\cr J}}}{2J+1}}\,
\langle{\bf n}|J 0 J M_J\rangle\,\biggr]\times
\nonumber\\[2mm]
&&\times\alpha_2{\cal N}_{12}^{(\alpha_1-\alpha_2)}(p)\,
R^{(\alpha_1\alpha_2)}_{J\pm11J}(p)\, ,
\nonumber\\[2mm]
\bigl[\tilde\Phi^{(\alpha_1\alpha_2)}_
{J{\tiny\pmatrix{0\cr1}}JM_J}({\bf p})\bigr]_{ba}&=&
\biggl[-\sqrt{\frac{{\footnotesize\pmatrix{J+1\cr J}}}{2J+1}}\,
\langle{\bf n}|J+1 1 J M_J\rangle
\pm\sqrt{\frac{{\footnotesize\pmatrix{J\cr J+1}}}{2J+1}}\,
\langle{\bf n}|J-1 1 J M_J\rangle\,\biggr]\times
\nonumber\\[2mm]
&&\times\alpha_1{\cal N}_{12}^{(-\alpha_1\alpha_2)}(p)\,
R^{(\alpha_1\alpha_2)}_{J{\tiny\pmatrix{0\cr 1}}J}(p)\, ,
\nonumber\\[2mm]
\bigl[\tilde\Phi^{(\alpha_1\alpha_2)}_
{J\pm11JM_J}({\bf p})\bigr]_{ba}&=&
\biggl[-\sqrt{\frac{{\footnotesize\pmatrix{J\cr J+1}}}{2J+1}}\,
\langle{\bf n}|J 1 J M_J\rangle
\mp\sqrt{\frac{{\footnotesize\pmatrix{J+1\cr J}}}{2J+1}}\,
\langle{\bf n}|J 0 J M_J\rangle\,\biggr]\times
\nonumber\\[2mm]
&&\times\alpha_1{\cal N}_{12}^{(-\alpha_1\alpha_2)}(p)\,
R^{(\alpha_1\alpha_2)}_{J\pm11J}(p)\, ,
\nonumber\\[2mm]
\bigl[\tilde\Phi^{(\alpha_1\alpha_2)}_
{J{\tiny\pmatrix{0\cr 1}}JM_J}({\bf p})\bigr]_{bb}&=&
\pm\langle{\bf n}|J{\footnotesize\pmatrix{0\cr 1}}JM_J\rangle\,
\alpha_1\alpha_2{\cal N}_{12}^{(-\alpha_1-\alpha_2)}(p)\,
R^{(\alpha_1\alpha_2)}_{J{\tiny\pmatrix{0\cr 1}}J}(p)\, ,
\nonumber\\[2mm]
\bigl[\tilde\Phi^{(\alpha_1\alpha_2)}_{J\pm11JM_J}({\bf p})\bigr]_{bb}&=&
\biggl[\pm \frac{1}{2J+1}\,\langle{\bf n}|J\pm 1 1 J M_J\rangle
-\frac{2\sqrt{J(J+1)}}{2J+1}\,\langle{\bf n}|J\mp 1 1 J M_J\rangle\biggr]\times
\nonumber\\[2mm]
&&\times\alpha_1\alpha_2{\cal N}_{12}^{(-\alpha_1-\alpha_2)}(p)\,
R^{(\alpha_1\alpha_2)}_{J\pm11J}(p)\, .
\en
With the use of these expressions, we can explicitly calculate the quantity
\eq\label{hat-phi}
\bigl[\hat{\tilde\Phi}^{(\alpha_1\alpha_2)}_{LSJM_J}({\bf p})\bigr]_{ij}=
\bigl[\tilde\Phi^{(\alpha_1\alpha_2)}_{LSJM_J}({\bf p})\bigr]_{ij}(-i\sigma_y)\, ,
\quad\quad i,j=a,b\, ,
\en
as an $2\times 2$ matrix in the fermion spin space. To this end, we explicitly introduce the fermion spin coordinates $\sigma_1$ and $\sigma_2$ 
($\sigma_i=\pm\frac{1}{2},~i=1,2$). Then, we have
\eq\label{spin-ex}
\langle{\bf n}|LSJM_J\rangle\equiv
\langle{\bf n}\sigma_1\sigma_2|LSJM_J\rangle
=\sum_{m_Lm_S}\langle LSm_Lm_S|JM_J\rangle\,
\langle\frac{1}{2}\frac{1}{2}\sigma_1\sigma_2|Sm_S\,
\langle{\bf n}|Lm_L\rangle\, .
\en
Further, with the account of the following relations
\eq\label{sigma_y}
\langle\frac{1}{2}\frac{1}{2}\sigma_1\sigma_2|00\rangle(-i\sigma_y)&=&
\frac{1}{2}\pmatrix{1&0\cr0&1}=\frac{1}{2}\, I\equiv\hat\varphi_0\, ,
\nonumber\\[2mm]
\langle\frac{1}{2}\frac{1}{2}\sigma_1\sigma_2|10\rangle(-i\sigma_y)&=&
\frac{1}{\sqrt{2}}\pmatrix{1&0\cr0&-1}=\frac{1}{\sqrt{2}}\, \sigma_z
\equiv\hat\varphi_{10}\, ,
\nonumber\\[2mm]
\langle\frac{1}{2}\frac{1}{2}\sigma_1\sigma_2|1\pm 1\rangle(-i\sigma_y)&=&
\pmatrix{0&{\tiny\pmatrix{-1\cr0}}\cr{\tiny\pmatrix{0\cr 1}}&0}
=\frac{1}{2}\, (\mp\sigma_x-i\sigma_y)=\frac{1}{\sqrt{2}}\,\sigma_\pm
\equiv\hat\varphi_{1\pm1}\, ,
\en
it follows that
\eq\label{hat-phi-not}
\langle{\bf n}|LSJM_J\rangle=\sum_{m_Lm_S}
\langle LSM_Lm_S|JM_J\rangle\,\langle{\bf n}|Lm_L\rangle\,
\hat\varphi_{Sm_S}\equiv(\langle{\bf n}|L\rangle\otimes\hat\varphi_S)^{JM_J}\, .
\en
For the quantity $\hat{\tilde\Phi}_{LSJM_J}({\bf p})=\sum_{\alpha_1\alpha_2}
\hat{\tilde\Phi}^{(\alpha_1\alpha_2)}_{LSJM_J}({\bf p})$ we obtain
\eq\label{hat-phi-all}
\bigl[\hat{\tilde\Phi}_{LSJM_J}({\bf p})\bigr]_{aa}&=&
(\langle{\bf n}|L\rangle\otimes\hat\varphi_S)^{JM_J}\sum_{\alpha_1\alpha_2}
{\cal N}_{12}^{(\alpha_1\alpha_2)}(p)R_{LSJ}^{(\alpha_1\alpha_2)}(p)\, ,
\nonumber\\[2mm]
\bigl[\hat{\tilde\Phi}_{J{\tiny\pmatrix{0\cr1}}JM_J}({\bf p})\bigr]_{ab}&=&
\biggl[\mp\sqrt{\frac{{\footnotesize\pmatrix{J\cr J+1}}}{2J+1}}\,
(\langle{\bf n}|J+1\rangle\otimes\hat\varphi_1)^{JM_J}
+\sqrt{\frac{{\footnotesize\pmatrix{J+1\cr J}}}{2J+1}}\,
(\langle{\bf n}|J-1\rangle\otimes\hat\varphi_1)^{JM_J}\biggr]\times
\nonumber\\[2mm]
&&\times\sum_{\alpha_1\alpha_2}\alpha_2{\cal N}_{12}^{(\alpha_1-\alpha_2)}(p)\,
R^{(\alpha_1\alpha_2)}_{J{\tiny\pmatrix{0\cr 1}}J}(p)\, ,
\nonumber\\[2mm]
\bigl[\hat{\tilde\Phi}_{J\pm11JM_J}({\bf p})\bigr]_{ab}&=&
\biggl[\sqrt{\frac{{\footnotesize\pmatrix{J\cr J+1}}}{2J+1}}\,
(\langle{\bf n}|J\rangle\otimes\hat\varphi_1)^{JM_J}
\mp\sqrt{\frac{{\footnotesize\pmatrix{J+1\cr J}}}{2J+1}}\,
(\langle{\bf n}|J\rangle\otimes\hat\varphi_0)^{JM_J}\biggr]\times
\nonumber\\[2mm]
&&\times\sum_{\alpha_1\alpha_2}\alpha_2{\cal N}_{12}^{(\alpha_1-\alpha_2)}(p)\,
R^{(\alpha_1\alpha_2)}_{J\pm11J}(p)\, ,
\nonumber\\[2mm]
\bigl[\hat{\tilde\Phi}_{J{\tiny\pmatrix{0\cr1}}JM_J}({\bf p})\bigr]_{ba}&=&
\biggl[-\sqrt{\frac{{\footnotesize\pmatrix{J+1\cr J}}}{2J+1}}\,
(\langle{\bf n}|J+1\rangle\otimes\hat\varphi_1)^{JM_J}
\pm\sqrt{\frac{{\footnotesize\pmatrix{J\cr J+1}}}{2J+1}}\,
(\langle{\bf n}|J-1\rangle\otimes\hat\varphi_1)^{JM_J}\biggr]\times
\nonumber\\[2mm]
&&\times\sum_{\alpha_1\alpha_2}\alpha_1{\cal N}_{12}^{(-\alpha_1\alpha_2)}(p)\,
R^{(\alpha_1\alpha_2)}_{J{\tiny\pmatrix{0\cr 1}}J}(p)\, ,
\nonumber\\[2mm]
\bigl[\hat{\tilde\Phi}_{J\pm11JM_J}({\bf p})\bigr]_{ba}&=&
\biggl[-\sqrt{\frac{{\footnotesize\pmatrix{J\cr J+1}}}{2J+1}}\,
(\langle{\bf n}|J\rangle\otimes\hat\varphi_1)^{JM_J}
\mp\sqrt{\frac{{\footnotesize\pmatrix{J+1\cr J}}}{2J+1}}\,
(\langle{\bf n}|J\rangle\otimes\hat\varphi_0)^{JM_J}\biggr]\times
\nonumber\\[2mm]
&&\times\sum_{\alpha_1\alpha_2}\alpha_1{\cal N}_{12}^{(-\alpha_1\alpha_2)}(p)\,
R^{(\alpha_1\alpha_2)}_{J\pm11J}(p)\, ,
\nonumber\\[2mm]
\bigl[\hat{\tilde\Phi}_{J{\tiny\pmatrix{0\cr 1}}JM_J}({\bf p})\bigr]_{bb}&=&
\biggl[\pm\biggl(\langle{\bf n}|J\rangle\otimes
\pmatrix{\hat\varphi_0\cr\hat\varphi_1}\biggr)^{JM_J}\biggr]
\sum_{\alpha_1\alpha_2}\alpha_1\alpha_2
{\cal N}_{12}^{(-\alpha_1-\alpha_2)}(p)
R^{(\alpha_1\alpha_2)}_{J{\tiny\pmatrix{0\cr 1}}J}(p)\, ,
\nonumber\\[2mm]
\bigl[\hat{\tilde\Phi}_{J\pm11JM_J}({\bf p})\bigr]_{bb}&=&
\biggl[\pm\frac{1}{2J+1}\,(\langle{\bf n}|J\pm1\rangle
\otimes\hat\varphi_1)^{JM_J}
-\frac{2\sqrt{J(J+1)}}{2J+1}\,(\langle{\bf n}|J\mp1\rangle
\otimes\hat\varphi_1)^{JM_J}\biggr]\times
\nonumber\\[2mm]
&&\times\sum_{\alpha_1\alpha_2}\alpha_1\alpha_2
{\cal N}_{12}^{(-\alpha_1-\alpha_2)}(p)
R^{(\alpha_1\alpha_2)}_{J\pm11J}(p)\, .
\en
In order to evaluate the constants $f_P$ and $f_V$, we need the bound-state
wave function of the $q\bar q$ state at ${\bf r}=0$
\eq\label{r=0}
&&\tilde\Psi_{LSJM_J}({\bf r}=0)\equiv
\tilde\Psi_{LSJM_J}({\bf r}=0,\sigma_1,\sigma_2)=
\int\frac{d^3{\bf p}}{(2\pi)^3}\,
\tilde\Psi_{LSJM_J}({\bf p})\equiv
\int\frac{d^3{\bf p}}{(2\pi)^3}\,
\tilde\Psi_{LSJM_J}({\bf p},\sigma_1\sigma_2)\, ,
\nonumber\\
&&
\en
where, according to Eq.~(\ref{Psi-C}), 
\eq\label{pw-Psi-C}
\tilde\Psi_{LSJM_J}({\bf p})=
\pmatrix{(\tilde\Psi_{LSJM_J}({\bf p}))_{ab} &
(\tilde\Psi_{LSJM_J}({\bf p}))_{aa}\cr
(\tilde\Psi_{LSJM_J}({\bf p}))_{bb} &
(\tilde\Psi_{LSJM_J}({\bf p}))_{ba}}\, .
\en

The decay constants $f_P$ and $f_V$ for the pseudoscalar ($L=S=J=0$)
and vector ($L=0,~S=J=1$) mesons, respectively, are given by~\cite{Royen}
\eq\label{fP-fV}
\delta_{\mu0}M_Bf_P&=&\sqrt{3}\,{\rm tr}\,
\bigl[\tilde\Psi_{0000}({\bf r}=0)\gamma^\mu(1-\gamma^5)\bigr]\, ,
\nonumber\\[2mm]
f_V(\lambda)&=&\sqrt{3}\,{\rm tr}\,\bigl[
\tilde\Psi_{011\lambda}({\bf r}=0)\gamma^\mu\bigr]\varepsilon_\mu(\lambda)\, ,
\quad\quad\lambda=\pm1,0\, ,
\en
where the factor $\sqrt{3}$ stems from the color part of the wave function, 
and $\varepsilon_\mu(\lambda)$ is the polarization vector of the vector 
meson~\cite{Helzen}
\eq\label{polarization}
\varepsilon_\mu(\lambda=\pm 1)=\mp\frac{1}{\sqrt{2}}\,(0,1,\pm i,0)\, ,
\quad\quad
\varepsilon_\mu(\lambda=0)=(0,0,0,1)\, ,\quad\quad
{\rm in~CM~frame}\, .
\en
Now, using the equations~(\ref{r=0})-(\ref{fP-fV}), we obtain
\eq\label{fP-fV-wf}
f_P&=&\frac{\sqrt{24\pi}}{M_B}\,
\int_0^\infty\frac{p^2dp}{(2\pi)^3}\,
\sum_{\alpha_1\alpha_2}\bigl[{\cal N}_{12}^{(\alpha_1\alpha_2)}(p)
-\alpha_1\alpha_2{\cal N}_{12}^{(-\alpha_1\alpha-2)}(p)\bigr]\,
R^{(\alpha_1\alpha_2)}_{000}(p)\, ,
\\[2mm] 
f_V(\lambda)&=&-\delta_{\lambda0}\sqrt{24\pi}\,
\int_0^\infty\frac{p^2dp}{(2\pi)^3}\,
\sum_{\alpha_1\alpha_2}\bigl[{\cal N}_{12}^{(\alpha_1\alpha_2)}(p)
+\frac{\alpha_1\alpha_2}{3}\, {\cal N}_{12}^{(-\alpha_1\alpha-2)}(p)\bigr]\,
R^{(\alpha_1\alpha_2)}_{011}(p)\equiv\delta_{\lambda0}f_V\, .
\nonumber
\en
The leptonic decay width of the vector mesons ($\rho^0,\omega,\Phi$) is given 
by
\eq\label{leptonic-width}
\Gamma(V\rightarrow e^+e^-)=4\pi\,\frac{\alpha_{eff}^2}{M_B^3}\,\,
\frac{1}{3}\,\sum_{\lambda=\pm1,0}|f_V(\lambda)|^2=
\frac{4\pi\alpha_{eff}^2|f_V|^2}{3M_B^3}\, ,
\en
where
\eq\label{alpha_eff}
\alpha_{eff}^2=\alpha^2\bar e_q^2\, ,\quad\quad
\bar e_q=e_q/e\, ,\quad\quad
\bar e_q^2=(\frac{1}{2},\frac{1}{18},\frac{1}{9})\, ,
\en
for $\rho^0,\omega,\Phi$ mesons, respectively. Here, $\bar e_q$
denotes the expectation value of the quark charge in the units of the
elementary charge $e$. 

In order to explain the reason, why the quantity $\bar e_q$ appears in
the expression (\ref{leptonic-width}), let us note that the leptonic
decay of the vector meson in the lowest order in $e$ is described by
the diagram depicted in Fig.~1. Taking into account the flavor
structure of the wave functions
\eq\label{flavour-structure}
\rho^0\sim\frac{1}{\sqrt{2}}\,(u\bar u-d\bar d)\, ,\quad\quad 
\omega\sim\frac{1}{\sqrt{2}}\,(u\bar u+d\bar d)\, ,\quad\quad 
\Phi\sim s\bar s\, ,
\en
we obtain, that the transition amplitudes of the vector mesons into
the photon are proportional to
\eq\label{V-gamma}
(\rho^0\rightarrow\gamma)\sim\frac{e}{\sqrt{2}}\, ,\quad\quad
(\omega\rightarrow\gamma)\sim\frac{e}{3\sqrt{2}}\, ,\quad\quad
(\Phi\rightarrow\gamma)\sim-\frac{e}{3}\, ,
\en
from which the Eq.~(\ref{alpha_eff}) follows directly.

Further, taking into account Eqs.~(\ref{basis}), (\ref{osc-NR-wf}) and 
(\ref{normalization-c}), we can express the quantity $f_P$ and the
leptonic decay width in terms of the dimensionless wave functions
$\bar R^{(\alpha_1\alpha_2)}_{LSJ}(z)$
\eq\label{dimensionless}
&&f_P=\frac{\sqrt{6}\bar p^{3/2}}{\pi\sqrt{M_B}}\,
\biggl|\sum_{\alpha_1\alpha_2}\int_0^\infty z^2dz\biggl[
{\cal N}_{12}^{(\alpha_1\alpha_2)}(\bar p,z)-\alpha_1\alpha_2
{\cal N}_{12}^{(-\alpha_1-\alpha_2)}(\bar p,z)\biggr]
\bar R^{(\alpha_1\alpha_2)}_{000}(z)\biggr|\, ,
\\[2mm]
&&\Gamma(V\rightarrow e^+e^-)=\frac{8\alpha_{eff}^2\bar p^3}{\pi M_B^2}\,
\biggl|\sum_{\alpha_1\alpha_2}\int_0^\infty z^2dz\biggl[
{\cal N}_{12}^{(\alpha_1\alpha_2)}(\bar p,z)+\frac{\alpha_1\alpha_2}{3}\,
{\cal N}_{12}^{(-\alpha_1-\alpha_2)}(\bar p,z)\biggr]
\bar R^{(\alpha_1\alpha_2)}_{011}(z)\biggr|^2\, ,
\nonumber
\en 
where the functions
\eq\label{bar-R}
\bar R^{(\alpha_1\alpha_2)}_{LSJ}(z)=\sum_{n=0}^\infty
C^{(\alpha_1\alpha_2)}_{LSJn}\, \bar R_{nL}(z)
\en
satisfy the normalization condition
\eq\label{norm-bar-R}
\sum_{\alpha_1\alpha_2}\int_0^\infty z^2dz f_{12}^{(\alpha_1\alpha_2)}
(M_B;\bar p,z)\,\biggl[\bar
R^{(\alpha_1\alpha_2)}_{LSJ}(z)\biggr]^2=1\, .
\en

Next, we consider the two-photon decays of the neutral
mesons.
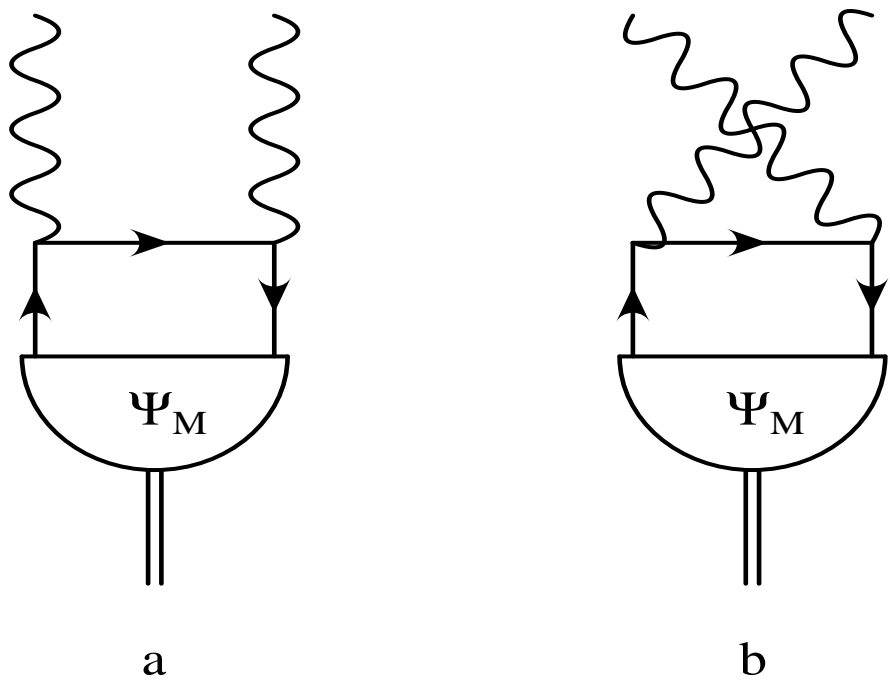
The amplitude of the two-photon decay of the $q\bar q$ bound state
with equal-mass quarks in the lowest order in the coupling constant
$e$ is given by the diagrams depicted in Fig.~2.
Th the CM frame where $P=(M_B,{\bf 0})$, this amplitude is equal
to~\cite{Resag-1} 
\eq\label{2-photon-amp}
&&T(\lambda_1\lambda_2)=i\sqrt{3}e_q^2\int\frac{d^4p}{(2\pi)^4}\,
{\rm tr}\,\biggl\{\Psi_{M_B}(p)\biggl[
\not\!\epsilon_1S(\frac{P}{2}+p-k_1)\not\!\epsilon_2
+\not\!\epsilon_2S(\frac{P}{2}+p-k_2)\not\!\epsilon_1
\biggr]\biggr\}\, ,
\en
where $\Psi_{M_B}(p)$ is the BS amplitude of the $q\bar q$ bound state
which satisfies Eq.~(\ref{BS-C}) and is written in the form~(\ref{Psi-C}). 
Further, $k_1=(\frac{M_B}{2},{\bf k})$ and $k_2=(\frac{M_B}{2},-{\bf k})$,
where ${\bf k}$ is the relative 3-momentum of photons in the CM frame
directed along the $z$-axis,
and $\epsilon_i\equiv\epsilon(\lambda_i)$ are the polarization vectors
for the emitted photons. Due to the fact that the emitted physical
photons are transversely polarized, one needs to consider only the
values $\lambda_i=\pm 1$ for which
\eq\label{epsilon_i}
\epsilon(\lambda_1=\pm1)=\mp\frac{1}{\sqrt{2}}\,(0,1,\pm i,0)\, ,\quad\quad
\epsilon(\lambda_2=\pm1)=\pm\frac{1}{\sqrt{2}}\,(0,1,\mp i,0)\, ,
\en
and
\eq\label{not-epsilon_i}
\not\!\epsilon(\lambda_1=\pm1)=\frac{1}{\sqrt{2}}\,(\pm\gamma_x+i\gamma_y)\, ,
\quad\quad
\not\!\epsilon(\lambda_2=\pm1)=\frac{1}{\sqrt{2}}\,(\mp\gamma_x+i\gamma_y)\, .
\en
Further, one may rewrite the expression, entering the integrand in 
Eq.~(\ref{2-photon-amp}) in the following manner
(Below, we follow the derivation given in Ref.~\cite{in-prep}.) 
\eq\label{f_M}
&&f_M(p,{\bf k})=
-i\not\!\epsilon_1S(\frac{P}{2}+p-k_1)\not\!\epsilon_2
-i\not\!\epsilon_2S(\frac{P}{2}+p-k_2)\not\!\epsilon_1
\\[2mm]
&&=\frac{a_{12}^{(+)}({\bf p}-{\bf k})}{p_0-w({\bf p}-{\bf k})+i0}
+\frac{a_{12}^{(-)}({\bf p}-{\bf k})}{p_0+w({\bf p}-{\bf k})-i0}
+\frac{a_{21}^{(+)}({\bf p}+{\bf k})}{p_0-w({\bf p}+{\bf k})+i0}
+\frac{a_{21}^{(-)}({\bf p}+{\bf k})}{p_0+w({\bf p}+{\bf k})-i0}\, ,
\nonumber
\en
where $w({\bf p}\pm{\bf k})=\sqrt{m^2+({\bf p}\pm{\bf k})^2}$, and
\eq\label{a-pm}
a_{12}^{(\alpha)}({\bf p}-{\bf k})
=\not\!\epsilon_1\Lambda^{(\alpha)}({\bf p}-{\bf k})\gamma_0\not\!\epsilon_2
\, ,\quad\quad
a_{21}^{(\alpha)}({\bf p}+{\bf k})
=\not\!\epsilon_2\Lambda^{(\alpha)}({\bf p}+{\bf k})\gamma_0\not\!\epsilon_1
\, .
\en
Note that, of course, the relation of the BS amplitude
$\Psi_{M_B}(p)$ and the 3D amplitude $\tilde\Psi_{M_B}({\bf p})$ 
is different in different versions of the 3D reduction.
In particular, in the Salpeter version,
\eq\label{ampl-SAL}
\Psi_{M_B}(p)=S(\frac{P}{2}+p)\Gamma({\bf p})S(-\frac{P}{2}+p)\, ,
\en
where, taking into account Eq.~(\ref{Psi-C}), we have
\eq\label{Gamma-SAL}
\Gamma({\bf p})=-i\int\frac{d^3{\bf p}'}{(2\pi)^3)}\, V({\bf p},{\bf
p}')\,
\tilde\Psi_{M_B}({\bf p}')\, ,\quad\quad
\tilde\Psi_{M_B}({\bf p}')=-i\pmatrix{
\tilde\phi_{ab}({\bf p}')\sigma_y &
\tilde\phi_{aa}({\bf p}')\sigma_y \cr
\tilde\phi_{bb}({\bf p}')\sigma_y &
\tilde\phi_{ba}({\bf p}')\sigma_y}\, .
\en
On the other hand, from Eq.~(\ref{ampl-SAL}) one may obtain
\eq\label{Psi_M}
\Psi_{M_B}(p)=
&-&\frac{\Gamma^{(+-)}({\bf p})}{M_B-2w+i0}\,\biggl(
\frac{1}{p_0-\frac{M_B}{2}+w-i0}-\frac{1}{p_0+\frac{M_B}{2}-w+i0}\biggr)
\nonumber\\[2mm]
&+&\frac{\Gamma^{(-+)}({\bf p})}{M_B+2w}\,\biggl(
\frac{1}{p_0-\frac{M_B}{2}+w-i0}-\frac{1}{p_0+\frac{M_B}{2}-w+i0}\biggr)
\nonumber\\[2mm]
&+&\frac{\Gamma^{(++)}({\bf p})}{M_B}\,\biggl(
\frac{1}{p_0-\frac{M_B}{2}-w+i0}-\frac{1}{p_0+\frac{M_B}{2}-w+i0}\biggr)
\nonumber\\[2mm]
&+&\frac{\Gamma^{(--)}({\bf p})}{M_B}\,\biggl(
\frac{1}{p_0-\frac{M_B}{2}+w-i0}-\frac{1}{p_0+\frac{M_B}{2}+w+i0}\biggr)\, ,
\en
where
\eq\label{Gamma_ab}
\Gamma^{(\alpha\beta)}({\bf p})=\Lambda^{(\alpha)}({\bf p})\gamma_0
\Gamma({\bf p})\gamma_0\Lambda^{(\beta)}(-{\bf p})\, .
\en
After integrating Eq.~(\ref{Psi_M}) over $p_0$, we obtain the Salpeter
equation for the equal-time amplitude  
\eq\label{SAL-ampl}
\tilde\Psi_{M_B}({\bf p})=-\frac{i\Gamma^{(+-)}({\bf p})}{M_B-2w+i0}+
\frac{i\Gamma^{(-+)}({\bf p})}{M_B+2w}\, .
\en
Now, substituting~(\ref{Psi_M}) into the expression of the two-photon
decay amplitude~(\ref{2-photon-amp}) and integrating over $p_0$, we
obtain: $T(\pm\mp)=0$ and
\eq\label{T_pm_pm}
&&T(\pm\pm)=i\sqrt{3}e_q^2\int\frac{d^3{\bf p}}{(2\pi)^3}\,\biggl\{
\frac{1}{\frac{1}{4}M_B^2-(w+w({\bf p}-{\bf k}))^2}\,
{\rm tr}\,\biggl[\frac{1}{2}\,(\Gamma^{(++)}({\bf p})-\Gamma^{(--)}({\bf p}))
(\gamma_0\mp\gamma_5\gamma_z)
\nonumber\\[2mm]
&&+\biggl(\frac{1}{2}\,(\Gamma^{(++)}({\bf p})-\Gamma^{(--)}({\bf p}))
+\biggl(\frac{M_B}{M_B+2w}\,\Gamma^{(-+)}({\bf p})
+i(\frac{M_B}{2}+w+w({\bf p}-{\bf k}))\biggr)\tilde\Psi_{M_B}({\bf p})\biggr)
\times
\nonumber\\[2mm]
&&\times\frac{1}{w({\bf p}-{\bf k})}\,((1\pm\gamma_5\gamma_0\gamma_z)m
+(\gamma_z\mp\gamma_5\gamma_0)(p_z-k))\biggr]
\nonumber\\[2mm]
&&+\frac{1}{\frac{1}{4}M_B^2-(w+w({\bf p}+{\bf k}))^2}\,
{\rm tr}\,\biggl[\frac{1}{2}\,(\Gamma^{(++)}({\bf p})-\Gamma^{(--)}({\bf p}))
(\gamma_0\mp\gamma_5\gamma_z)
\nonumber\\[2mm]
&&+\biggl(\frac{1}{2}\,(\Gamma^{(++)}({\bf p})+\Gamma^{(--)}({\bf p}))
+\biggl(\frac{M_B}{M_B+2w}\,\Gamma^{(-+)}({\bf p})
+i(\frac{M_B}{2}+w+w({\bf p}+{\bf k}))\biggr)\tilde\Psi_{M_B}({\bf p})\biggr)
\times
\nonumber\\[2mm]
&&\times\frac{1}{w({\bf p}+{\bf k})}\,((1\mp\gamma_5\gamma_0\gamma_z)m
+(\gamma_z\mp\gamma_5\gamma_0)(p_z+k))\biggr]\biggr\}\, .
\en
For the further transformation of this expression, one may use the
fact that the static potential $V({\bf p},{\bf p}')$ has the Lorentz
structure given by Eq.~(\ref{V-static}). Then,
\eq\label{Psi_M-Lorentz}    
&&V({\bf p},{\bf p}')\tilde\Psi_{M_B}({\bf p}')=
V_{\rm OG}({\bf p}-{\bf p}')\gamma_0\tilde\Psi_{M_B}({\bf p}')\gamma_0
+V_{\rm C}({\bf p}-{\bf p}')\biggl(x\gamma_0\tilde\Psi_{M_B}({\bf p}')\gamma_0
\nonumber\\[2mm]
&&+(1-x)\tilde\Psi_{M_B}({\bf p}')\biggr)
+V_{\rm T}({\bf p}-{\bf p}')4\hat g(({\rm tr}\,\tilde\Psi_{M_B}({\bf p}'))+
\gamma_5{\rm tr}\,(\tilde\Psi_{M_B}({\bf p}')\gamma_5))\, .
\en
From this, one can directly obtain 
\eq\label{Gamma_ab-Lorentz}
\Gamma^{(\alpha\beta)}({\bf p})&=&-i\Lambda^{(\alpha)}({\bf p})
\int\frac{d^3{\bf p}'}{(2\pi)^3}\,
\biggl[ V_{\rm OG}({\bf p}-{\bf p}')
\pmatrix{\hat{\tilde{\phi}}_{ab}({\bf p}') &
\hat{\tilde{\phi}}_{aa}({\bf p}') \cr
\hat{\tilde{\phi}}_{bb}({\bf p}') &
\hat{\tilde{\phi}}_{ba}({\bf p}') }
\nonumber\\[2mm]
&+&V_{\rm C}({\bf p}-{\bf p}')
\pmatrix{\hat{\tilde{\phi}}_{ab}({\bf p}') &
(2x-1)\hat{\tilde{\phi}}_{aa}({\bf p}') \cr
(2x-1)\hat{\tilde{\phi}}_{bb}({\bf p}') &
\hat{\tilde{\phi}}_{ba}({\bf p}') }
\\[2mm]
&+&V_{\rm T}({\bf p}-{\bf p}')4\hat g
\pmatrix{{\rm tr}\,(\hat{\tilde{\phi}}_{ab}({\bf p}')
                   +\hat{\tilde{\phi}}_{ba}({\bf p}')) &
-{\rm tr}\,(\hat{\tilde{\phi}}_{aa}({\bf p}')
                   +\hat{\tilde{\phi}}_{bb}({\bf p}')) \cr
-{\rm tr}\,(\hat{\tilde{\phi}}_{aa}({\bf p}')
                   +\hat{\tilde{\phi}}_{bb}({\bf p}')) &
{\rm tr}\,(\hat{\tilde{\phi}}_{ab}({\bf p}')
                   +\hat{\tilde{\phi}}_{ba}({\bf p}'))}\biggr]\,
\Lambda^{(\beta)}(-{\bf p})\, ,
\nonumber
\en
where
\eq\label{components-spin}
\hat{\tilde{\phi}}_{\alpha\beta}({\bf p})=
-i\tilde\phi_{\alpha\beta}({\bf p})\sigma_y
\en
are the components of the meson amplitude in the spin space.
After performing the partial-wave decomposition of these amplitudes,
the expression for the quantity $T(\lambda_1\lambda_2)$ takes the
form (note that we have replaced t'Hooft interaction by its
regularized version).

\vspace*{.3cm}

{\bf $~^1S_0$ state:}

\eq\label{1S0}
&&T(\pm\pm)=\pm\frac{\sqrt{3}e_q^2}{(\sqrt{2}\pi)^5}\,\int_0^\infty\frac{p^2dp}
{w^2}\biggl\{\int {p'}^2dp'\biggl[\biggl(
\frac{2mw}{M_B(M_B+2w)}\, J_0(M_B;p)+\frac{2mw}{M_Bp}\, I_1(M_B;p)
\nonumber\\[2mm]
&&+\frac{2mw^2}{M_B^2p}\, I_2(M_B;p)\biggl)V_{\rm C}^1(p,p')R_{000(ab+ba)}(p')
-(2x-1)\biggl((\frac{p}{M_B}+\frac{m^2}{p(M_B+2w)})J_0(M_B;p)
\nonumber\\[2mm]
&&+\frac{2w}{M_B}\, I_1(M_B;p)+\frac{2w^2}{M_B^2}\, I_2(M_B;p)\biggl)
V_{\rm C}^0(p,p')R_{000(aa-bb)}(p')
\nonumber\\[2mm]
&&+(2x-1)\,\frac{mw}{p(M_B+2w)}\, J_0(M_B;p)\, V_{\rm C}^0(p,p') 
R_{000(aa+bb)}(p')\biggr]+\biggl[ x=1,V_{\rm C}\rightarrow V_{\rm OG}\biggr]
\nonumber\\[2mm]
&&+4\hat g\biggl[\frac{2mw}{p(M_B+2w)}\, J_0(M_B;p) 
V_{{\rm T},reg}^0(p,p';\Lambda) R_{000(aa-bb)}(p')\biggr]\biggr)
\nonumber\\[2mm]
&&+\frac{4w^2}{M_B}\biggl[-\frac{m}{p}I_3(M_B;p)R_{000(ab+ba)}(p)
+(\frac{M_B}{2p}\,
I_3^0(M_B;p)+I_3(M_B;p))R_{000(aa-bb)}(p)\biggr]\biggr\}\, ,
\en

\vspace*{.3cm}

{\bf $~^3P_0$ state:}

\eq\label{3P0}
&&T(\pm\pm)=\pm\frac{\sqrt{3}e_q^2}{(\sqrt{2}\pi)^5}\,\int_0^\infty\frac{p^2dp}
{w^2}\biggl\{\int {p'}^2dp'\biggl[\biggl(
\frac{2m}{M_B}(\frac{m^2}{pM_B}+\frac{p}{M_B+2w})J_0(M_B;p)
\nonumber\\[2mm]
&&+\frac{2mw}{M_Bp}\, I_1^0(M_B;p)
+\frac{2mw}{M_B(M_B+2w)}\, I_2(M_B;p)+\frac{mp}{M_B^2}\, I_2^2(M_B;p)
\biggr) V_{\rm C}^0(p,p') R_{110(ab+ba)}(p')
\nonumber\\[2mm]
&&-(2x-1)\biggl(\frac{4m^2w}{M_B^2(M_B+2w)}\, J_0(M_B;p)
+\frac{2w}{M_B}\, I_1^0(M_B;p)
+(\frac{p}{M_B}+\frac{m^2}{p(M_B+2w)})I_2(M_B;p)
\nonumber\\[2mm]
&&+2(\frac{p^2}{M_B^2}+\frac{m^2}{M_B(M_B+2w)})I_2^2(M_B;p)\biggr)
V_{\rm C}^1(p,p')R_{110(aa-bb)}(p')-(2x-1)\times
\nonumber\\[2mm]
&&\times\biggl(\frac{M_B}{M_B+2w}\,\frac{2mw}{M_B^2}\,
(J_0(M_B;p)-I_2^2(M_B;p))-\frac{mw}{M_Bp}\, I_2(M_B;p)\biggr)
V_{\rm C}^1(p,p')R_{110(aa+bb)}(p')\biggr]
\nonumber\\[2mm]
&&+\biggl[ x=1,V_{\rm C}\rightarrow V_{\rm OG}\biggr]
+8\hat g\biggr[ 4\biggl(\frac{mw}{M_Bp}\, I_1^0(M_B;p)
+\frac{m^3}{M_B^2p}\, J_0(M_B;p)
+\frac{m}{2M_B}\, I_2(M_B;p)
\nonumber\\[2mm]
&&+\frac{mp}{M_B^2}\, I_2^2(M_B;p)\biggr)
+\frac{4M_B}{M_B+2w}\biggl(
\frac{mp}{M_B^2}\, J_0(M_B;p)
-\frac{m}{M_B}\, I_2(M_B;p)
-\frac{mp}{M_B^2}\, I_2^2(M_B;p)\biggr)\biggr]\times
\nonumber\\[2mm]
&&\times V_{{\rm T},reg}^0(p,p';\Lambda)R_{110(ab+ba)}(p')\biggl)
+\frac{4w^2}{M_B}\biggl[
-\frac{m}{p}\, I_3^0(M_B;p)R_{110(ab+ba)}(p)
\nonumber\\[2mm]
&&+(\frac{M_B}{2p}I_3(M_B;p)-I_3^2(M_B;p))R_{110(aa-bb)}(p)\biggr]\biggr\}\, ,
\en
where
\eq\label{notations}
R_{LSJ(aa+bb)}(p)&=&R_{LSJ}^{(++)}(p)+R_{LSJ}^{(--)}(p)\, ,
\nonumber\\[2mm]
R_{LSJ(aa-bb)}(p)&=&\frac{m}{w}(R_{LSJ}^{(++)}(p)-R_{LSJ}^{(--)}(p))
+\frac{p}{w}(R_{LSJ}^{(+-)}(p)+R_{LSJ}^{(-+)}(p))\, ,
\nonumber\\[2mm]
R_{LSJ(ab+ba)}(p)&=&\frac{p}{w}(R_{LSJ}^{(++)}(p)-R_{LSJ}^{(--)}(p))
+\frac{m}{w}(R_{LSJ}^{(+-)}(p)+R_{LSJ}^{(-+)}(p))\, ,
\en
and
\eq\label{JI-s}
J_0&=&\ln\frac{2(w+w_+)-M_B}{2(w+w_+)+M_B}
-\ln\frac{2(w+w_-)-M_B}{2(w+w_-)+M_B}\, ,
\nonumber\\[2mm]
J&=&\ln\frac{2w(w+w_+)+M_Bp}{2w(w+w_-)-M_Bp}\, ,
\nonumber\\[2mm]
I_1^0&=&\frac{1}{2}\, J-\frac{w}{M_B}\, J_0\, ,
\nonumber\\[2mm]
I_1&=&1-\frac{4w}{w_++w_-}+\frac{w^2}{M_Bp}\, J -\frac{w}{2p}\, J_0\, ,
\nonumber\\[2mm]
I_2&=&\frac{2M_B}{w_++w_-}-\frac{w}{p}\, J\, ,
\nonumber\\[2mm]
I_2^2&=&-\frac{2w}{p}+\frac{2}{3}\, 
\frac{5w^2-\frac{1}{4}\, M_B^2+w_+w_-}{p(w_++w_-)}+\frac{w^2}{p^2}\,
J_0\, ,
\nonumber\\[2mm]
I_3^0&=&\ln\frac{2(w+w_+)-M_B}{2(w+w_-)-M_B}\, ,
\nonumber\\[2mm]
I_3&=&1+\frac{M_B-2w}{w_++w_-}-\frac{w}{p}\, I_3^0\, ,
\nonumber\\[2mm]
I_3^2&=&\frac{w}{p}+\frac{(M_B-2w)(w^2+\frac{1}{4}\, M_B^2-w_+w_-+3M_Bw)}
{3M_Bp(w_++w_-)}+\frac{w^2}{p^2}\, I_3^0\, ,
\en
with $w_\pm=\sqrt{w^2+\frac{1}{4}M_B^2\pm M_Bp}$.

Now, we consider other versions of the 3D equations. Since we
consider the equal-mass case, the Gross equation can not be used.
For this reason, we shall restrict ourselves to study of two-photon 
decay processes in CJ and MNK versions. In these versions, there
exists a
relation between 4D and 3D free Green functions given by
Eqs.~(\ref{CJ-G0-3-4}) and (\ref{MNK-3D-4D}). 
This relation can be immediately translated into the relation between
the 4D and 3D wave functions
\eq\label{4-3-wf}
\Psi_{M_B}(p)=2\pi i\delta(p_0)\,\tilde\Psi_{M_B}({\bf p})\, ,
\en
where for the MNK version the equality $p_0^+=0$ holds for the
equal-mass case. As to the MW, version, here the relation between
$\Psi_{M_B}(p)$ and $\tilde\Psi_{M_B}({\bf p})$ 
does not exist due to the definition of 
$\tilde G_0^{\rm MW}(M_B,{\bf p})$~(\ref{G0-MW}). For the above reasons,
below we restrict ourselves to the CJ and MNK versions
only. Substituting the expression~(\ref{4-3-wf}) 
into Eq.~(\ref{2-photon-amp}), with an account of (\ref{f_M}) one obtains
\eq\label{Tl1l2}
T(\lambda_1\lambda_2)=i\sqrt{3}e_q^2\int\frac{d^3{\bf p}}{(2\pi)^3}\,
{\rm tr}\,\biggl\{\tilde\Psi_{M_B}({\bf p})&&\biggl[
\frac{a_{12}^{(+)}({\bf p}-{\bf k})}{w({\bf p}-{\bf k})}
-\frac{a_{12}^{(-)}({\bf p}-{\bf k})}{w({\bf p}-{\bf k})}
\nonumber\\[2mm]
&&+\frac{a_{21}^{(+)}({\bf p}+{\bf k})}{w({\bf p}+{\bf k})}
-\frac{a_{21}^{(-)}({\bf p}+{\bf k})}{w({\bf p}+{\bf k})}
\biggr]\biggr\}\, .
\en
From this equation one readily obtains
\eq\label{T-pm}
T(\pm,\pm)=\pm i\sqrt{3}e_q^2\int\frac{d^3{\bf p}}{(2\pi)^3}\,
{\rm tr}\,\biggl\{\tilde\Psi_{M_B}({\bf p})&&\biggl[
\frac{(1\pm\gamma_5\gamma_0\gamma_z)m+(\gamma_z\mp\gamma_5\gamma_0)(p_z-k)}
{w^2({\bf p}-{\bf k})}
\nonumber\\[2mm]
&&+\frac{(1\mp\gamma_5\gamma_0\gamma_z)m+(\gamma_z\pm\gamma_5\gamma_0)(p_z+k)}
{w^2({\bf p}+{\bf k})}
\biggr]\biggr\}\, .
\en
Substituting $\tilde\Psi_{M_B}({\bf p})$ in the matrix form
given by Eq.~(\ref{pw-Psi-C}), we finally obtain for the CJ and MNK
versions

\vspace*{.3cm}

{\bf $^1S_0$ state:}

\eq\label{Tl1l2-1S0}
T(\pm\pm)=\pm ie_q^2\frac{2}{(2\pi)^{5/2}}\,
\int_0^\infty pdp&&\biggl[
\biggl(-2+\frac{w^2+\frac{1}{4}M_B^2}{M_Bp}\,
\tilde J(M_B;p)\biggr)\frac{m}{M_B}\, R_{000(ab+ba)}(p)
\nonumber\\[2mm]
&&+\biggl(\frac{2p}{M_B}-\frac{w^2-\frac{1}{4}M_B^2}{2M_B^2}\,
\tilde J(M_B;p)\biggr) R_{000(aa-bb)}(p)\biggr]\, ,
\en

\vspace*{.3cm}

{\bf $^3P_0$ state:}

\eq\label{Tl1l2-3P0}
T(\pm\pm)=-i\sqrt{3}e_q^2\frac{2}{(2\pi)^{5/2}}\,
&&\int_0^\infty pdp\biggl[
\tilde J(M_B;p)\frac{m}{M_B}\,R_{110(ab+ba)}(p)
\nonumber\\[2mm]
&&+2\biggl(\frac{w^2-\frac{1}{4}M_B^2}{M_B^2}-
\frac{w^4-\frac{1}{16}M_B^4}{2M_B^3p}\, \tilde J(M_B;p)\biggr)
R_{110(aa-bb)}(p)\biggr]\, ,
\en
where
\eq\label{tilde-J}
\tilde J(M_B;p)=\ln\,\frac{w^2+\frac{1}{4}M_B^2+M_Bp}
{w^2+\frac{1}{4}M_B^2-M_Bp}\, .
\en
It is important to note that in the Salpeter version the two-photon
decay amplitude depends on the potential both directly and indirectly,
through the radial wave functions, whereas in CJ and MNK versions this
dependence enters only through the radial wave functions.

For a given meson, the two-photon decay amplitude can be rewritten as
\eq\label{meson-2}
T(\lambda_1\lambda_2)=e^2\tilde e_{q,eff}^2\sqrt{3}
\tilde T(\lambda_1\lambda_2;LSJM_J)\, .
\en
The decay width is given by
\eq\label{2-photon-decay}
\Gamma({\rm meson}\rightarrow\gamma\gamma)=3\pi\,
\frac{\alpha^2}{M_B}\,\frac{1}{2(2J+1)}\,
\sum_{\lambda_1\lambda_2M_J}
\biggl|\tilde e_{q,eff}^2
\tilde T(\lambda_1\lambda_2;LSJM_J)\biggr|^2\, ,
\en
where $\tilde e_{q,eff}^2$ depends on the choice of the meson flavor
wave function. If this function has a simple form $q\bar q$ then
$\tilde e_{q,eff}^2=\tilde e_q^2$. However, if the meson wave function
is made up of different flavor states $\alpha q_1\bar q_1
+\beta q_2\bar q_2$, the expression for $\tilde e_{q,eff}^2$ is more 
complicated. Consider as an example calculation of this factor for
$\pi^0$ and $\eta_n$ states. The flavor structure of the wave
functions is given by
\eq\label{fl-struc}
\pi^0\sim\frac{1}{\sqrt{2}}\,(u\bar u-d\bar d)\, ,\quad\quad
\eta_n\sim\frac{1}{\sqrt{2}}\,(u\bar u+d\bar d)\, .
\en
It follows then straightforwardly that 
$\tilde e_{q,eff}^2=\frac{1}{3\sqrt{2}}$ and
$\tilde e_{q,eff}^2=\frac{5}{9\sqrt{2}}$ for $\pi^0$ and $\eta_n$
states, respectively. Further, the decay amplitudes for the physical 
$\eta$ and $\eta'$ mesons are the linear superposition of the ones 
corresponding to $\eta_n$ and $\eta_s\sim s\bar s$ states.

Note that the two-photon decays of $\pi^0,~\eta,~\eta'$ mesons was
also studied in the NJL model, taking into account the relativistic
confinement and the t'Hooft interaction~\cite{Shakin-2}.

\vspace*{.5cm}

{\em Acknowledgments.} The author thanks T. Babutsidze and A. Rusetsky
for useful discussions.

\end{document}